\documentclass[showpacs,preprintnumbers,amsmath,amssymb]{revtex4}

\usepackage{graphicx}
\usepackage{dcolumn}
\usepackage{bm}

\newcommand{\bq}{\begin{equation}}
\newcommand{\eq}{\end{equation}}
\newcommand{\bqa}{\begin{eqnarray}}
\newcommand{\eqa}{\end{eqnarray}}
\newcommand{\nn}{\nonumber \\}

\def\be     {\begin{equation}}
\def\ee     {\end{equation}}
\def\bea        {\begin{eqnarray}}
\def\eea        {\end{eqnarray}}
\def\bnn    {\begin{eqnarray*}}
\def\enn    {\end{eqnarray*}}

\begin{document}

\title{Boltzmann-equation approach to anomalous transport in a Weyl metal}
\author{Ki-Seok Kim$^{1,2}$, Heon-Jung Kim$^{3}$, and M. Sasaki$^{4}$}
\affiliation{ $^{1}$Department of Physics, POSTECH, Pohang, Gyeongbuk 790-784, Korea \\ $^{2}$Institute of Edge of
Theoretical Science (IES), Hogil Kim Memorial building 5th floor, POSTECH, Pohang, Gyeongbuk 790-784, Korea \\
$^{3}$Department of Physics, College of Natural Science, Daegu University, Gyeongbuk 712-714, Korea \\
$^{4}$Department of Physics, Faculty of Science, Yamagata University, Kojirakawa, Yamagata 990-8560, Japan }
\date{\today}

\begin{abstract}
Weyl metal is regarded as a platform toward interacting topological states of matter, where its topological structure gives rise to anomalous transport phenomena, referred to as chiral magnetic effect and ``negative" magneto-resistivity, the origin of which is chiral anomaly. Recently, the negative magneto-resistivity has been observed with the signature of weak anti-localization at $x = 3 \sim 4 ~ \%$ in Bi$_{1-x}$Sb$_{x}$, where magnetic field is applied in parallel with electric field ($\bm{E} \parallel \bm{B}$). Based on the Boltzmann-equation approach, we find the negative magneto-resistivity in the presence of weak anti-localization. An essential ingredient is to introduce the topological structure of chiral anomaly into the Boltzmann-equation approach, resorting to semi-classical equations of motion with Berry curvature.
%
%This theoretical framework allows us to predict an unconventional ``normal" Hall effect contributed from the Fermi surface of each Weyl cone in the case of $\bm{E} \parallel \bm{B}$, which originates from the topological $\bm{E}\cdot\bm{B}$ term. Unfortunately, such an anomalous ``normal" Hall effect turns out to be canceled when all contributions of both Fermi surfaces are summed.
\end{abstract}

\maketitle

\section{Introduction}

It is the endless mission of condensed matter physics to search novel quantum states of matter. Since the discovery of the concept of topological insulators \cite{Kane_Mele_Model,Zhang_Bernevig_Model,3D_TI_Z2index_FuKane,3D_TI_Z2index_Balents,3D_TI_Z2index_Roy}, the topological structure of quantum matter lies at the center of research for novel quantum matter. Recalling that electron correlations have been playing an essential role in emergent phenomena of quantum matter, a research on the interplay between topology and interaction seems to drive the direction of condensed matter physics at present.

Weyl metal is regarded as a platform toward interacting topological states of matter. Its metallicity allows us to introduce electron correlations via doping, giving rise to possible instabilities of their Fermi surfaces. Its topological structure is encoded by chiral anomaly \cite{Chiral_Anomaly}, responsible for anomalous transport phenomena referred to as chiral magnetic effect \cite{Chiral_Magnetic_Effect1,Chiral_Magnetic_Effect2,CME_Kubo1,CME_Kubo2,Chiral_Magnetic_Effect3,Chiral_Magnetic_Effect4} and negative magneto-resistivity \cite{Negative_LMR1,Negative_LMR2,Negative_LMR3}. In this respect we would like to call an effective theory for Weyl metal topological Landau's Fermi-liquid theory \cite{Haldane} and topological Landau-Ginzburg framework for phase transitions. This direction of research is expected to lead a branch of condensed matter physics.

First of all, the characteristic feature of Weyl metal originates from its band structure. Let's start from the band structure of a topological insulator, described by an effective Dirac Hamiltonian in momentum space \cite{TI_Band_Structure}
\bqa && Z = \int D \psi_{\sigma\tau}(\boldsymbol{k}) \exp \Bigl\{ - \int_{0}^{\beta} d \tau \int \frac{d^{3}
\boldsymbol{k}}{(2\pi)^{3}} \psi_{\sigma\tau}^{\dagger}(\boldsymbol{k}) \Bigl(
(\partial_{\tau} - \mu) \boldsymbol{I}_{\sigma\sigma'} \otimes \boldsymbol{I}_{\tau\tau'} + v
\boldsymbol{k} \cdot \boldsymbol{\sigma}_{\sigma\sigma'} \otimes
\boldsymbol{\tau}_{\tau\tau'}^{z} + m(|\boldsymbol{k}|) \boldsymbol{I}_{\sigma\sigma'} \otimes
\boldsymbol{\tau}_{\tau\tau'}^{x} \Bigr) \psi_{\sigma'\tau'}(\boldsymbol{k}) \Bigr\} . \nonumber \eqa
Here, $\psi_{\sigma\tau}(\boldsymbol{k})$ represents a four-component Dirac spinor, where $\sigma$ and $\tau$ are spin and chiral indexes, respectively.
$\bm{\sigma}_{\sigma\sigma'}$ and $\bm{\tau}_{\tau\tau'}$ are Pauli matrices acting on spin and ``orbital" spaces. The relativistic dispersion is represented in the chiral basis, where each eigen value of $\boldsymbol{\tau}_{\tau\tau'}^{z}$ expresses either $+$ or $-$ chirality, respectively. The mass term can be formulated as $m(|\boldsymbol{k}|) = m - \rho |\bm{k}|^{2}$, where $\mbox{sgn}(m) \mbox{sgn}(\rho) > 0$ corresponds to a topological insulating state while $\mbox{sgn}(m) \mbox{sgn}(\rho) < 0$ does to a normal band insulating phase. $\mu$ is the chemical potential, controlled by doping. One may regard that this simplified effective model can be derived from a realistic band structure in Bi$_{1-x}$Sb$_{x}$, describing dynamics of electrons near the $\bm{L}-$point in momentum space.

It has been demonstrated that the mass gap can be tuned to vanish at $x = 3 \sim 4 ~ \%$ in Bi$_{1-x}$Sb$_{x}$, allowing us to reach the critical point between the topological and band insulating phases \cite{BiSb1,BiSb2,BiSb3}. It is straightforward to show that this gapless Dirac spectrum splits into two Weyl points, breaking time reversal symmetry, for example, applying magnetic field into the gapless semi-conductor
\bqa && H_{TRB} = g_{\psi} \psi_{\sigma\tau}^{\dagger}(\boldsymbol{k}) (\boldsymbol{H} \cdot \boldsymbol{\sigma}_{\sigma\sigma'} \otimes \boldsymbol{I}_{\tau\tau'}) \psi_{\sigma'\tau'}(\boldsymbol{k}) , \nonumber \eqa where $g_{\psi}$ is the Land\'{e} g-factor. The band touching point $(0,0,0)$ of the Dirac spectrum shifts into $(0,0, g_{\psi} H / v)$ and $(0,0,- g_{\psi} H / v)$ for each chirality along the direction of magnetic field, given by
\bqa && E_{\bm{k}} + \mu = \pm \sqrt{v^{2}[k_{x}^{2} + k_{y}^{2}] + [g_{\psi} H \pm v k_{z}]^{2}} . \nonumber \eqa Now, each spectrum is described by a two-component Weyl spinor with a definite chirality, referred to as Weyl metal \cite{Weyl_Metal1,Weyl_Metal2,Weyl_Metal3}. One can also find this type of spectrum, breaking inversion symmetry instead of time reversal symmetry.

An interesting feature of Weyl metal results from the fact that each Weyl point can be identified with a magnetic monopole in momentum space. In other words, each $\pm$ magnetic charge becomes ``polarized" in momentum space, applying magnetic field. As a result, a Fermi arc, which connects such magnetic monopole and anti-monopole pairs in the bulk, appears on the surface state \cite{Weyl_Metal2}, exactly analogous to the Weyl point on the surface state of a topological insulator, where each Fermi point of the Fermi arc corresponds to the Weyl point of the case of the topological insulator. Unfortunately, this spectroscopic fingerprint has not been observed yet.

In our opinion the characteristic feature of Weyl metal is beyond the Berry curvature given by the band structure. A cautious person may point out that the band structure of Weyl metal is essentially the same as that of graphene except for the existence of the Fermi arc, where the $+$ chirality Weyl spectrum at the $\bm{K}$ point and the $-$ chirality Weyl spectrum at the $-\bm{K}$ point allow us to call graphene two-dimensional Weyl metal \cite{Review_Graphene}. However, there is critical one difference between Weyl metal and graphene. Weyl electrons in the paired Weyl points are not independent in Weyl metal while they have ``nothing" to do with each other in graphene. It is true that Weyl points in graphene can be regarded as a pair of Weyl points with opposite chirality according to the no-go theorem by Nielsen and Ninomiya \cite{NoGoTheorem1,NoGoTheorem2}. In addition, they can be shifted and merged into one Dirac point, applying effective ``magnetic" fields to couple with the pseudo-spin of graphene. However, there does not exist such an anomaly relation between the pair of Weyl points in graphene, which means that currents are conserved separately for each Weyl cone in contrast with the case of Weyl metal as long as inter-valley scattering can be neglected. A crucial different aspect between two and three dimensions is that the irreducible representation of the Lorentz group is a four-component Dirac spinor in three dimensions while it is a two-component Weyl spinor in two dimensions. As a result, the pair of Weyl points originates from the Dirac point in three dimensions, where such a pair of Weyl points is ``connected" with each other through the Dirac sea. On the other hand, each Weyl point of the pair exists ``independently" in two dimensions. Chiral anomaly is the key feature of Weyl metal.

Suppose QED$_{4}$ with a topological $\bm{E}\cdot\bm{B}$ term,
\bqa && Z = \int D \psi \exp\Bigl[ - \int_{0}^{\beta} d \tau \int d^{3} \bm{r} \Bigl\{ \bar{\psi} \Bigl( i \gamma^{\mu} [\partial_{\mu} + i e A_{\mu}] + \mu \gamma^{0} \Bigr) \psi - \frac{1}{4} F_{\mu\nu} F^{\mu\nu} + \theta \frac{e^{2}}{16 \pi^{2}} \epsilon^{\mu\nu\rho\delta} F_{\mu\nu} F_{\rho\delta} \Bigr\} \Bigr] , \nonumber \eqa where $\psi$ is a four-component Dirac spinor and $F_{\mu\nu} = \partial_{\mu} A_{\nu} - \partial_{\nu} A_{\mu}$ is electromagnetic field-strength tensor with electromagnetic field $A_{\mu}$. Chiral anomaly means that the chiral symmetry preserved in the classical level is not respected any more in the quantum level due to the presence of special types of quantum fluctuations, given by the triangle diagram \cite{Peskin_Schroeder}. As a result, the associated chiral current, the right-handed chiral current minus the left-handed chiral current, is not conserved in the quantum field theory, described by
\bqa && \partial_{\mu} (\bar{\psi} \gamma^{\mu} \gamma^{5} \psi) = - \frac{e^{2}}{16 \pi^{2}} \epsilon^{\mu\nu\rho\delta} F_{\mu\nu} F_{\rho\delta} , \nonumber \eqa where $\bar{\psi} \gamma^{\mu} \gamma^{5} \psi = \bar{\psi}_{+} \gamma^{\mu} \psi_{+} - \bar{\psi}_{-} \gamma^{\mu} \psi_{-}$ is the chiral current with the $\pm$ chiral charge. Resorting to this chiral anomaly, we can rewrite the above expression as follows \cite{Jho_Kim},
\bqa && Z = \int D \psi \exp\Bigl[ - \int_{0}^{\beta} d \tau \int d^{3} \bm{r} \Bigl\{ \bar{\psi} \Bigl( i \gamma^{\mu} [\partial_{\mu} + i e A_{\mu} + i c_{\mu} \gamma^{5} ] + \mu \gamma^{0} \Bigr) \psi - \frac{1}{4} F_{\mu\nu} F^{\mu\nu} \Bigr\} \Bigr] , \nonumber \eqa where the chiral gauge field is given by \bqa && c_{\mu} = \partial_{\mu} \theta . \nonumber \eqa Representing the Dirac gamma matrix in the chiral basis, it is straightforward to identify the chiral gauge field with the applied magnetic field in the previous effective model Hamiltonian. In other words, the Dirac point splits into one paired Weyl points, the origin of which is chiral anomaly with breaking either time reversal symmetry or inversion symmetry, encoded in $\partial_{\mu} \theta \not= 0$.

It turns out that the chiral anomaly is responsible for anomalous transport phenomena in Weyl metal \cite{Chiral_Magnetic_Effect1,Chiral_Magnetic_Effect2,Chiral_Magnetic_Effect3,Chiral_Magnetic_Effect4,Negative_LMR1,Negative_LMR2,Negative_LMR3}. Recently, we could measure the negative magneto-resistivity with the signature of weak anti-localization \cite{Negative_LMR3}, regarded as one transport fingerprint with the chiral magnetic effect. As discussed before, a Weyl metallic state is expected to appear applying magnetic field into the Dirac metal, believed to be realized at the topological critical point in Bi$_{1-x}$Sb$_{x}$ with $x = 3 \sim 4 \%$. The negative magneto-resistivity has been observed only when electric currents are driven along the direction of magnetic field, saying $\bm{E} \parallel \bm{B}$, where $\bm{E}$ is electric field. Recalling that electron correlations would be negligible in this metallic phase, this strong anisotropy in magneto-resistivity has been attributed to the topological $\bm{E} \cdot \bm{B}$ term.

In this study we discuss the origin of the negative magneto-resistivity based on the Boltzmann-equation approach. An idea is to introduce the topological structure of chiral anomaly into the Boltzmann-equation approach \cite{Negative_LMR2}, resorting to semi-classical equations of motion which encode the information of Berry curvature \cite{Semiclassical_Eqs1,Semiclassical_Eqs2}. In addition to the introduction of chiral anomaly with the Berry curvature, we incorporate weak anti-localization quantum corrections into the negative magneto-resistivity phenomenologically \cite{WAL_Boltzmann}, the original expression of which is to consider the Drude conductivity for each Weyl fermion \cite{Negative_LMR2}. This theoretical framework allows us to investigate another type of anomalous Hall effect in the case of $\bm{E} \parallel \bm{B}$, which differs from the ``conventional" anomalous Hall effect \cite{AHE_Kubo1,AHE_Kubo2} in the case of $\bm{E} \perp \bm{B}$. The former is based on the presence of the topological $\bm{E}\cdot\bm{B}$ term, which plays the role of an additional force in dynamics of Weyl fermions beyond the conventional Lorentz force, while the latter originates from the appearance of an anomalous velocity due to the Berry curvature itself. It turns out that such an anomalous Hall effect does not exist in contrast with the claim of Ref. \cite{Negative_LMR3}.

\section{Review on the Boltzmann-equation approach for Weyl metal}

We would like to review the topological aspect of Weyl metal based on the Boltzmann-equation approach \cite{Negative_LMR2} for general readership. First, we re-derive the hydrodynamic equation from Boltzmann equation, where the $\bm{E}\cdot\bm{B}$ term encoded by the semi-classical equation-of-motion approach breaks the conservation law for the chiral current. Second, we re-derive the chiral magnetic effect from Boltzmann equation, where a subtle issue on the chiral magnetic effect, not transparent in the Boltzmann-equation approach, is also touched.

\subsection{Chiral anomaly}

A phenomenological Boltzmann equation is
\bqa && \Bigl( \frac{\partial}{\partial t} + \dot{\bm r} \cdot \bm{\nabla}_{\bm{r}} + \dot{\bm p} \cdot \bm{\nabla}_{\bm{p}} \Bigr) f(\bm{p};\bm{r},t) = I_{coll}[f(\bm{p};\bm{r},t)] , \eqa which can be derived based on the Schwinger-Keldysh formulation, where $f(\bm{p};\bm{r},t)$ is the distribution function with the conjugate momentum $\bm{p}$ of the relative coordinate and the center of mass coordinate ($\bm{r}$, $t$) in the Wigner transformation of the lesser Green's function \cite{Mahan_Book}. The right-hand side represents a collision term, incorporating electron correlations and impurity scattering effects.

An essential idea is to introduce the information of the topological structure into the Boltzmann equation via the semi-classical equation-of-motion approach \cite{Negative_LMR2}, given by
\bqa && \boldsymbol{\dot{r}} = \frac{\partial \epsilon_{\boldsymbol{p}}}{\partial \boldsymbol{p}} + \boldsymbol{\dot{p}} \times \boldsymbol{\Omega}_{\boldsymbol{p}}, \nn && \boldsymbol{\dot{p}} = e \boldsymbol{E} + \frac{e}{c} \boldsymbol{\dot{r}} \times \boldsymbol{B} , \eqa where $\boldsymbol{\Omega}_{\boldsymbol{p}} = \bm{\nabla}_{\bm{p}} \times \bm{A}_{\bm{p}}$ is the Berry curvature and $\bm{A}_{\bm{p}} = i \langle u_{\bm{p}} | \bm{\nabla}_{\bm{p}} u_{\bm{p}} \rangle$ is the Berry connection with the Bloch's eigen state $| u_{\bm{p}} \rangle$ \cite{Semiclassical_Eqs1,Semiclassical_Eqs2}. It is straightforward to find the solution of these semi-classical equations of motion, given by
\bqa && \boldsymbol{\dot{r}} = \Bigl( 1 + \frac{e}{c} \boldsymbol{B} \cdot \boldsymbol{\Omega}_{\boldsymbol{p}} \Bigr)^{-1} \Bigl\{ \boldsymbol{v}_{\boldsymbol{p}} + e \boldsymbol{E} \times \boldsymbol{\Omega}_{\boldsymbol{p}} + \frac{e}{c} \boldsymbol{\Omega}_{\boldsymbol{p}} \cdot \boldsymbol{v}_{\boldsymbol{p}} \boldsymbol{B} \Bigr\} , \nn && \boldsymbol{\dot{p}} = \Bigl( 1 + \frac{e}{c} \boldsymbol{B} \cdot \boldsymbol{\Omega}_{\boldsymbol{p}} \Bigr)^{-1} \Bigl\{ e \boldsymbol{E} + \frac{e}{c} \boldsymbol{v}_{\boldsymbol{p}} \times \boldsymbol{B} + \frac{e^{2}}{c} (\boldsymbol{E} \cdot \boldsymbol{B}) \boldsymbol{\Omega}_{\boldsymbol{p}} \Bigr\} . \eqa Here, $\boldsymbol{v}_{\boldsymbol{p}} = \bm{\nabla}_{\bm{p}} \epsilon_{\boldsymbol{p}}$ with a band structure $\epsilon_{\boldsymbol{p}}$. We would like to point out that this band structure needs not be linear-in-momentum strictly. It is important that a Fermi surface encloses a Weyl cone while the structure of the Fermi surface needs not be limited to the Weyl-band structure. Focusing on dynamics of electrons on the Fermi surface, it doesn't look much different from that on a ``normal" Fermi surface. However, these electrons experience effects of both Berry curvature and chiral anomaly on the Fermi surface as long as the Fermi surface encloses the Weyl-type cone, regarded to be the characteristic feature toward a topological Fermi-liquid theory \cite{Haldane}. As will be discussed below, the second term in the $\bm{\dot{r}}$ equation results in the anomalous Hall effect \cite{AHE_Kubo1,AHE_Kubo2} given by the Berry curvature \cite{Semiclassical_Eqs1,Semiclassical_Eqs2} and the third term gives rise to the chiral magnetic effect \cite{Chiral_Magnetic_Effect1,Chiral_Magnetic_Effect2,CME_Kubo1,CME_Kubo2,Chiral_Magnetic_Effect3,Chiral_Magnetic_Effect4} while the last term in the $\bm{\dot{p}}$ equation is the source of chiral anomaly, responsible for the negative magneto-resistivity \cite{Negative_LMR1,Negative_LMR2,Negative_LMR3}.

%\bqa && \boldsymbol{\Omega}_{\boldsymbol{p}} = \bm{\nabla}_{\bm{p}} \times \bm{A}_{\bm{p}} , ~~~~~ \bm{A}_{\bm{p}} = i \langle u_{\bm{p}} | \bm{\nabla}_{\bm{p}} u_{\bm{p}} \rangle \eqa

Applying this idea into Weyl metal, we can write down an effective theory in the Boltzmann-equation approach,
\bqa && \frac{\partial f^{+}(\bm{p};\bm{r},t)}{\partial t} + \Bigl( 1 + \frac{e}{c}
\boldsymbol{B} \cdot \boldsymbol{\Omega}^{+}_{\boldsymbol{p}} \Bigr)^{-1} \Bigl\{ \boldsymbol{v}_{\boldsymbol{p}} + e \boldsymbol{E} \times \boldsymbol{\Omega}^{+}_{\boldsymbol{p}} + \frac{e}{c} \boldsymbol{\Omega}^{+}_{\boldsymbol{p}} \cdot
\boldsymbol{v}_{\boldsymbol{p}} \boldsymbol{B} \Bigr\} \cdot \bm{\nabla}_{\bm{r}} f^{+}(\bm{p};\bm{r},t) \nn && + \Bigl( 1 + \frac{e}{c} \boldsymbol{B} \cdot \boldsymbol{\Omega}^{+}_{\boldsymbol{p}} \Bigr)^{-1} \Bigl\{ e \boldsymbol{E} + \frac{e}{c} \boldsymbol{v}_{\boldsymbol{p}} \times \boldsymbol{B} + \frac{e^{2}}{c} (\boldsymbol{E} \cdot
\boldsymbol{B}) \boldsymbol{\Omega}^{+}_{\boldsymbol{p}} \Bigr\} \cdot \bm{\nabla}_{\bm{p}} f^{+}(\bm{p};\bm{r},t)
= I_{coll}^{+}[f^{+}(\bm{p};\bm{r},t), f^{-}(\bm{p};\bm{r},t)] , \nn && \frac{\partial f^{-}(\bm{p};\bm{r},t)}{\partial t} + \Bigl( 1 + \frac{e}{c}
\boldsymbol{B} \cdot \boldsymbol{\Omega}^{-}_{\boldsymbol{p}} \Bigr)^{-1} \Bigl\{ \boldsymbol{v}_{\boldsymbol{p}} + e \boldsymbol{E} \times \boldsymbol{\Omega}^{-}_{\boldsymbol{p}} + \frac{e}{c} \boldsymbol{\Omega}^{-}_{\boldsymbol{p}} \cdot
\boldsymbol{v}_{\boldsymbol{p}} \boldsymbol{B} \Bigr\} \cdot \bm{\nabla}_{\bm{r}} f^{-}(\bm{p};\bm{r},t) \nn && + \Bigl( 1 + \frac{e}{c} \boldsymbol{B} \cdot \boldsymbol{\Omega}^{-}_{\boldsymbol{p}} \Bigr)^{-1} \Bigl\{ e \boldsymbol{E} + \frac{e}{c} \boldsymbol{v}_{\boldsymbol{p}} \times \boldsymbol{B} + \frac{e^{2}}{c} (\boldsymbol{E} \cdot
\boldsymbol{B}) \boldsymbol{\Omega}^{-}_{\boldsymbol{p}} \Bigr\} \cdot \bm{\nabla}_{\bm{p}} f^{-}(\bm{p};\bm{r},t)
= I_{coll}^{-}[f^{-}(\bm{p};\bm{r},t), f^{+}(\bm{p};\bm{r},t)] , \eqa where the $\pm$ superscript represents the $\pm$ chirality. In other words, we write down the Boltzmann equation near each Weyl point, where inter Weyl-point scattering is introduced into the collision term. The information of a magnetic monopole and anti-monopole pair is encoded by the opposite sign of magnetic charges, \bqa && \bm{\nabla}_{\bm{p}} \cdot \bm{\Omega}_{\bm{p}}^{+} = \delta^{(3)}(\bm{p} - g_{\psi} \bm{B}), ~~~~~ \bm{\nabla}_{\bm{p}} \cdot \bm{\Omega}_{\bm{p}}^{-} = - \delta^{(3)}(\bm{p} + g_{\psi} \bm{B}) , \eqa where $2 g_{\psi} \bm{B}$ corresponds to the distance between the paired Weyl points, as discussed in the introduction.

It is not that difficult to derive the hydrodynamic equation from the Boltzmann equation, resorting to the coarse graining procedure in the momentum space \cite{Negative_LMR2}. As a result, we reach the following expression
\bqa && \frac{\partial N^{\pm}}{\partial t} + \bm{\nabla}_{\bm{r}} \cdot \bm{j}^{\pm} = k^{\pm} \frac{e^{2}}{4\pi^{2}} \bm{E} \cdot \bm{B} , \eqa where $N^{\pm} = \int_{-\infty}^{\infty} d \epsilon \rho^{\pm}(\epsilon) f^{\pm}(\epsilon;\bm{r},t)$ and $\bm{j}^{\pm} = \int \frac{d^{3} \bm{p}}{(2\pi)^{3}} \Bigl( 1 + \frac{e}{c} \boldsymbol{B} \cdot \boldsymbol{\Omega}^{\pm}_{\boldsymbol{p}} \Bigr) \bm{\dot{r}} f^{\pm}(\bm{p};\bm{r},t) = \int \frac{d^{3} \bm{p}}{(2\pi)^{3}} \Bigl\{ \boldsymbol{v}_{\boldsymbol{p}} + e \boldsymbol{E} \times \boldsymbol{\Omega}^{\pm}_{\boldsymbol{p}} + \frac{e}{c} \boldsymbol{\Omega}^{\pm}_{\boldsymbol{p}} \cdot \boldsymbol{v}_{\boldsymbol{p}} \boldsymbol{B} \Bigr\} f^{\pm}(\bm{p};\bm{r},t)$ are the density with the density of states $\rho^{\pm}(\epsilon) = \int \frac{d^{3} \bm{p}}{(2\pi)^{3}} \Bigl( 1 + \frac{e}{c} \boldsymbol{B} \cdot \boldsymbol{\Omega}^{\pm}_{\boldsymbol{p}} \Bigr) \delta(\epsilon_{\bm{p}} - \epsilon)$ and current, respectively, around each Weyl point. $k^{\pm} = \frac{1}{2\pi} \int d \bm{S}_{\bm{p}} \cdot \boldsymbol{\Omega}^{\pm}_{\boldsymbol{p}} = \pm 1$ is a magnetic charge at each Weyl point. It is clear that the current conservation law around each Weyl point breaks down due to the $\bm{E}\cdot\bm{B}$ term, introduced by the semi-classical equation of motion, while the collision term does not play the role of either a source or sink. Interestingly, the $+$ chiral charge plays the role of a source in this hydrodynamic equation while the $-$ chiral charge does that of a sink. As a result, the total current is conserved, given by
%
%\bqa && N^{\pm} = \int_{-\infty}^{\infty} d \epsilon \rho^{\pm}(\epsilon) f^{\pm}(\epsilon;\bm{r},t) , ~~~~~ \bm{j}^{\pm} = \int \frac{d^{3} \bm{p}}{(2\pi)^{3}} \Bigl\{ \boldsymbol{v}_{\boldsymbol{p}} + e \boldsymbol{E} \times %\boldsymbol{\Omega}^{\pm}_{\boldsymbol{p}} + \frac{e}{c} \boldsymbol{\Omega}^{\pm}_{\boldsymbol{p}} \cdot \boldsymbol{v}_{\boldsymbol{p}} \boldsymbol{B} \Bigr\} f^{\pm}(\bm{p};\bm{r},t) \eqa
%
%\bqa && \rho^{\pm}(\epsilon) = \int \frac{d^{3} \bm{p}}{(2\pi)^{3}} \Bigl( 1 + \frac{e}{c} \boldsymbol{B} \cdot \boldsymbol{\Omega}^{\pm}_{\boldsymbol{p}} \Bigr) \delta(\epsilon_{\bm{p}} - \epsilon) \eqa
%
\bqa && \frac{\partial (N^{+} + N^{-})}{\partial t} + \bm{\nabla}_{\bm{r}} \cdot (\bm{j}^{+}+\bm{j}^{-}) = 0  \eqa while the chiral current is not, described by
\bqa && \frac{\partial (N^{+} - N^{-})}{\partial t} + \bm{\nabla}_{\bm{r}} \cdot (\bm{j}^{+}-\bm{j}^{-}) = \frac{e^{2}}{2\pi^{2}} \bm{E} \cdot \bm{B} . \eqa This is the chiral anomaly.

We would like to emphasize that this Boltzmann-equation approach is applicable only when the chemical potential lies away from the Weyl point, forming a pair of Fermi surfaces. When the chemical potential touches the Weyl point, we should re-derive the Boltzmann equation from QED$_{4}$. The distribution function in this relativistic Boltzmann equation will be expressed as a $4 \times 4$ matrix since the lesser Green's function consists of the four-component spinor. An interesting and fundamental problem is ``Taking the non-relativistic limit from the matrix Boltzmann equation when the chemical potential lies above the Weyl point, can we reproduce the present Boltzmann-equation framework, where effects of other components except for the Fermi-surface component are ``integrated out" or ``coarse grained", giving rise to such contributions as semi-classical equations of motion?" This research will give a formal basis to the present phenomenological Boltzmann-equation approach. Recently, the Boltzmann-equation framework has been derived from QED$_{4}$, based on the introduction of the Wigner function to satisfy a quantum kinetic equation \cite{QBE_WM1,QBE_WM2,QBE_WM3}. These derivations imply that Lorentz symmetry, gauge symmetry, and quantum mechanics are important ingredients for the existence of chiral anomaly.

\subsection{Chiral magnetic effect}

There is an interesting transport signature in Weyl metal, referred to as chiral magnetic effect \cite{Chiral_Magnetic_Effect1,Chiral_Magnetic_Effect2,CME_Kubo1,CME_Kubo2,Chiral_Magnetic_Effect3,Chiral_Magnetic_Effect4}, proposed to appear in ``equilibrium", i.e., $\bm{E} = 0$. The total electric current is given by
\bqa && \bm{j} = \bm{j}^{+}+\bm{j}^{-} = \frac{e}{c} \int \frac{d^{3} \bm{p}}{(2\pi)^{3}} \Bigl\{ (\boldsymbol{\Omega}^{+}_{\boldsymbol{p}} \cdot \boldsymbol{v}_{\boldsymbol{p}}) \boldsymbol{B} f^{+}(\bm{p};\bm{r},t) + (\boldsymbol{\Omega}^{-}_{\boldsymbol{p}} \cdot \boldsymbol{v}_{\boldsymbol{p}}) \boldsymbol{B} f^{-}(\bm{p};\bm{r},t) \Bigr\}  \eqa when $\bm{E} = 0$, where the distribution function is an equilibrium one. Considering
$\bm{\Omega}_{\bm{p}}^{+} \approx - \bm{\Omega}_{\bm{p}}^{-} = \bm{\Omega}_{\bm{p}}$ from Eq. (5), we obtain \bqa && \bm{j} = \frac{e}{c} \int \frac{d^{3} \bm{p}}{(2\pi)^{3}} (\boldsymbol{\Omega}_{\boldsymbol{p}} \cdot
\boldsymbol{v}_{\boldsymbol{p}}) [f^{+}(\bm{p};\bm{r},t) - f^{-}(\bm{p};\bm{r},t)] \boldsymbol{B} = \mathcal{C} (e/c) (\mu_{+} - \mu_{-}) \bm{B} \eqa at zero temperature, where the constant coefficient is given by $\mathcal{C} \approx \int \frac{d^{3} \bm{p}}{(2\pi)^{3}} (\boldsymbol{\Omega}_{\boldsymbol{p}} \cdot \boldsymbol{v}_{\boldsymbol{p}})$. In spite of zero electric field, electric currents turn out to flow along the direction of magnetic field in Weyl metal as long as the ``chiral" chemical potential ($\mu_{+} - \mu_{-}$) is finite. Although this transport phenomenon is beyond our imagination, there is a subtle issue, not transparent in the Boltzmann-equation approach. First of all, it looks counter-intuitive that the electric current can flow in equilibrium since applying infinitesimal electric field to this current state gives rise to power generation proportional to $\bm{j} \cdot \bm{E}$, where the Weyl metallic state is compressible. This implies that energy can be extracted out from the ground state, causing a paradox in the definition of the ground state \cite{Chiral_Magnetic_Effect4,Weyl_Metal3}. It has been discussed that the chiral magnetic effect depends on the limiting procedure for the transferred momentum and frequency \cite{Weyl_Metal3}. If one sets frequency to be zero first, then the system is in equilibrium and the chiral magnetic effect turns out to vanish. On the other hand, if one chooses the limit of $\bm{q} = 0$ first, then the system is away from equilibrium and the chiral magnetic effect does not vanish, given by the above expression. Unfortunately, this subtle issue is hidden in this Boltzmann-equation approach.

\section{Anomalous transport in Weyl metal}

Another transport fingerprint is the negative magneto-resistivity which occurs only when the electric current is driven along the direction of the paired Weyl points, originating from the topological $\bm{E} \cdot \bm{B}$ term. As discussed in the introduction, our recent experiments measured this anomalous transport phenomenon only when the electric field is applied in parallel with the magnetic field \cite{Negative_LMR1,Negative_LMR2,Negative_LMR3}. In addition to this weird longitudinal transport, we also observed weak anti-localization corrections in the magneto-resistivity for both cases of $\bm{E} \parallel \bm{B}$ and $\bm{E} \perp \bm{B}$ \cite{Negative_LMR3}. In this respect we need to introduce such quantum corrections into the Boltzmann-equation approach. Unfortunately, this derivation has not been performed systematically as far as we know. Instead, there is somewhat a phenomenological approach, where the introduction of a non-local scattering term into the collision integral reproduces the weak anti-localization correction in the electrical resistivity \cite{WAL_Boltzmann}.

\subsection{Review on the Boltzmann-equation approach with weak localization or weak anti-localization quantum corrections}

We start from an extended Boltzmann equation \bqa && \Bigl( \frac{\partial}{\partial t} + \dot{\bm r} \cdot \bm{\nabla}_{\bm{r}} + \dot{\bm p} \cdot \bm{\nabla}_{\bm{p}} \Bigr) f(\bm{p};\bm{r},t)
= - \Gamma_{imp} [f(\bm{p};\bm{r},t) - f_{eq}(\bm{p})] - \int_{-\infty}^{t} d t' \alpha(t-t') [f(-\bm{p};\bm{r},t') - f_{eq}(\bm{p})] . \eqa The collision part consists of two scattering contributions. The first is an elastic impurity-scattering term in the relaxation-time approximation, where $\Gamma_{imp}^{-1} = (2 \pi n_{I} |V_{imp}|^{2} N_{F})^{-1}$ with an impurity concentration $n_{I}$ and its potential strength $V_{imp}$ corresponds to the mean free time, the time scale between events of impurity scattering \cite{Mahan_Book}. The second is a weak-localization (weak anti-localization) term, expressed in a non-local way, which originates from multiple impurity scattering. $\alpha(t-t') = \pm \frac{\Gamma_{imp}}{\pi N_{F}} \int \frac{d^{3} \bm{q}}{(2\pi)^{3}} \exp\Bigl\{ - (D \bm{q}^{2} + \tau_{\phi}^{-1}) (t-t') \Bigr\}$ may be regarded as the diffusion kernel, which becomes more familiar, performing Fourier transformation as follows \cite{WAL_Boltzmann} \bqa && \alpha(\nu) = \pm \int_{-\infty}^{t} d t' e^{i \nu(t-t')} \alpha(t-t') = \pm \frac{\Gamma_{imp}}{\pi N_{F}} \int \frac{d^{3} \bm{q}}{(2\pi)^{3}} \frac{1}{D \bm{q}^{2} - i \nu + \tau_{\phi}^{-1}} , \eqa where the sign of $+$ ($-$) represents the weak localization (weak anti-localization). $D$ is the diffusion coefficient and $N_{F}$ is the density of states at the Fermi energy. This expression is supplemented by the upper cut-off in the momentum integral, given by the reciprocal of the mean free path $\Gamma_{imp}/v_{F}$ with the Fermi velocity $v_{F}$, and $\tau_{\phi}$ corresponds to the lower cut-off, identified with the phase-coherence lifetime.
%
%\bqa && \Gamma_{imp} = 2 \pi n_{I} |V_{imp}|^{2} N_{F} \eqa
%

Let's confirm that this extended Boltzmann equation recovers the well-known weak-localization (weak anti-localization) formula. For simplicity, we consider a simple metal without the contribution of Berry curvature $\boldsymbol{\Omega}_{\boldsymbol{p}} = 0$. Performing the Fourier transformation of $f(\bm{p};t) = \int_{-\infty}^{\infty} d \nu e^{- i \nu t} f(\bm{p};\nu)$, the Boltzmann equation reads
\bqa && \Bigl\{ - i \nu + \Bigl( e \boldsymbol{E} + \frac{e}{c} \boldsymbol{v}_{\boldsymbol{p}} \times \boldsymbol{B} \Bigr) \cdot \bm{\nabla}_{\bm{p}} \Bigr\} f(\bm{p};\nu) = - \Gamma_{imp} [f(\bm{p};\nu) - f_{eq}(\bm{p})] - \alpha(\nu) [f(-\bm{p};\nu) - f_{eq}(\bm{p})] . \eqa

Consider the standard setup $\bm{E} = E \hat{\bm x}$ and $\bm{B} = B \hat{\bm z}$ for magneto-resistivity and Hall measurements. Then, the Boltzmann equation is written as follows
\bqa && \Bigl\{\Gamma_{imp} - i \nu + \alpha(\nu) - \frac{e B}{c} \Bigl( v_{x}(\bm{p}) \frac{\partial}{\partial p_{y}} - v_{y}(\bm{p}) \frac{\partial}{\partial p_{x}} \Bigr) \Bigr\} f(\bm{p};\nu) = [\Gamma_{imp} + \alpha(\nu)] f_{eq}(\bm{p}) - e E \frac{\partial}{\partial p_{x}} f_{eq} (\bm{p}) \eqa in the linear response regime, where $f(-\bm{p};\nu)$ is replaced with $f(\bm{p};\nu)$ in the weak-localization (weak anti-localization) term. This interchange is allowed when both time reversal symmetry and inversion symmetry are preserved. Although the time reversal symmetry is not respected by the applied magnetic field, we resort to their approximate correspondence. Instead, the lower cut-off of the phase-coherence time is given by a function of the external magnetic field. Then, it is straightforward to show that the resulting weak-localization (weak anti-localization) correction in the magneto-resistivity coincides with its well-known expression.

This Boltzmann equation leads us to propose the following ansatz for the distribution function
\bqa && f(\bm{p};\nu) = \frac{\Gamma_{imp} + \alpha(\nu)}{\Gamma_{imp} - i \nu + \alpha(\nu)} f_{eq}(\bm{p}) - \frac{1}{\Gamma_{imp} - i \nu + \alpha(\nu)} e E \frac{\partial}{\partial p_{x}} f_{eq} (\bm{p}) + \Bigl( - \frac{\partial}{\partial \epsilon} f_{eq} (\epsilon) \Bigr) \bm{v}_{\bm{p}} \cdot \bm{\Lambda}(\bm{p};\nu) , \eqa where $\bm{\Lambda}(\bm{p};\nu)$ corresponds to a correction that arises from the presence of magnetic field.

Inserting this expression into the Boltzmann equation, we obtain
\bqa && \frac{e B}{m c} \frac{e E v_{y}(\bm{p})}{\Gamma_{imp} - i \nu + \alpha(\nu)} - \frac{e B}{m c} \Bigl( v_{x}(\bm{p}) \Lambda_{y}(\bm{p};\nu) - v_{y}(\bm{p}) \Lambda_{x}(\bm{p};\nu) \Bigr) + [\Gamma_{imp} - i \nu + \alpha(\nu)] \bm{v}_{\bm{p}} \cdot \bm{\Lambda}(\bm{p};\nu) = 0 , \eqa where $m$ is a band mass of an electron on the Fermi surface which encloses a Weyl point. It is defined from the Fermi velocity of $\bm{v}_{F} = \frac{\bm{p}_{F}}{m}$, where $\bm{p}_{F}$ is a Fermi momentum. Since this equation should be satisfied for any values of velocity, we find \bqa && \Lambda_{z}(\bm{p};\nu) = 0 . \eqa

Introducing $V(\bm{p}) = v_{x}(\bm{p}) + i v_{y}(\bm{p})$ and $\Lambda(\bm{p};\nu) = \Lambda_{x}(\bm{p};\nu) - i \Lambda_{y}(\bm{p};\nu)$ into the above expression, we reach
\bqa && \Re \Bigl\{ - i \frac{e E \omega_{c}}{\Gamma_{imp} - i \nu + \alpha(\nu)} V(\bm{p}) + [\Gamma_{imp} - i \nu - i \omega_{c} + \alpha(\nu)] V(\bm{p}) \Lambda(\bm{p};\nu)  \Bigr\} = 0 , \eqa where $\Re$ represents a real part and $\omega_{c} = \frac{e B}{m c}$ is the cyclotron frequency. It is straightforward to solve this equation, the solution of which is given by
\bqa && \Lambda_{x}(\bm{p};\nu) = - e E \frac{\omega_{c} (2 \nu + \omega_{c}) [\Gamma_{imp} + \alpha(\nu)]}{\Bigl( [\Gamma_{imp} + \alpha(\nu)]^{2} - \nu (\nu + \omega_{c}) \Bigr)^{2} + (2 \nu + \omega_{c})^{2} [\Gamma_{imp} + \alpha(\nu)]^{2}} \eqa and \bqa && \Lambda_{y}(\bm{p};\nu) = - e E \frac{\omega_{c} \Bigl( [\Gamma_{imp} + \alpha(\nu)]^{2} - \nu (\nu + \omega_{c}) \Bigr)}{\Bigl( [\Gamma_{imp} + \alpha(\nu)]^{2} - \nu (\nu + \omega_{c}) \Bigr)^{2} + (2 \nu + \omega_{c})^{2} [\Gamma_{imp} + \alpha(\nu)]^{2}} . \eqa Then, we reach the following expression for the distribution function,
\bqa && f(\bm{p};\nu) = \frac{\Gamma_{imp} + \alpha(\nu)}{\Gamma_{imp} - i \nu + \alpha(\nu)} f_{eq}(\bm{p}) + e E v_{x}(\bm{p}) \Bigl( - \frac{\partial}{\partial \epsilon} f_{eq} (\epsilon) \Bigr) \frac{1}{\Gamma_{imp} - i \nu + \alpha(\nu)} \nn && - e E v_{x}(\bm{p}) \Bigl( - \frac{\partial}{\partial \epsilon} f_{eq} (\epsilon) \Bigr) \frac{\omega_{c} (2 \nu + \omega_{c}) [\Gamma_{imp} + \alpha(\nu)]}{\Bigl( [\Gamma_{imp} + \alpha(\nu)]^{2} - \nu (\nu + \omega_{c}) \Bigr)^{2} + (2 \nu + \omega_{c})^{2} [\Gamma_{imp} + \alpha(\nu)]^{2}} \nn && - e E v_{y}(\bm{p}) \Bigl( - \frac{\partial}{\partial \epsilon} f_{eq} (\epsilon) \Bigr) \frac{\omega_{c} \Bigl( [\Gamma_{imp} + \alpha(\nu)]^{2} - \nu (\nu + \omega_{c}) \Bigr)}{\Bigl( [\Gamma_{imp} + \alpha(\nu)]^{2} - \nu (\nu + \omega_{c}) \Bigr)^{2} + (2 \nu + \omega_{c})^{2} [\Gamma_{imp} + \alpha(\nu)]^{2}} . \eqa

Recalling the current formula $\bm{j}(\nu) = e \int \frac{d^{3} \bm{p}}{(2\pi)^{3}} \dot{\bm r} f(\bm{p};\nu)$, we find an optical magneto-conductivity and optical Hall coefficient, given by
\bqa && \sigma_{xx}(\nu) = \frac{n e^{2}}{m} \Bigl\{ \frac{1}{\Gamma_{imp} - i \nu + \alpha(\nu)} - \frac{\omega_{c} (2 \nu + \omega_{c}) [\Gamma_{imp} + \alpha(\nu)]}{\Bigl( [\Gamma_{imp} + \alpha(\nu)]^{2} - \nu (\nu + \omega_{c}) \Bigr)^{2} + (2 \nu + \omega_{c})^{2} [\Gamma_{imp} + \alpha(\nu)]^{2}} \Bigr\}  \eqa and
\bqa && \sigma_{yx}(\nu) = - \frac{n e^{2}}{m} \frac{\omega_{c} \Bigl( [\Gamma_{imp} + \alpha(\nu)]^{2} - \nu (\nu + \omega_{c}) \Bigr)}{\Bigl( [\Gamma_{imp} + \alpha(\nu)]^{2} - \nu (\nu + \omega_{c}) \Bigr)^{2} + (2 \nu + \omega_{c})^{2} [\Gamma_{imp} + \alpha(\nu)]^{2}} , \eqa respectively, where \bqa && \frac{n}{m} = \int \frac{d^{3} \bm{p}}{(2\pi)^{3}} [v_{x}(\bm{p})]^{2} \Bigl( - \frac{\partial}{\partial \epsilon} f_{eq} (\epsilon) \Bigr) \eqa with an electron density $n$ contributed from a Fermi surface and its band mass $m$.

The dc-limit of the above formulae is given by
\bqa && \sigma_{xx} = \sigma_{imp} \frac{1 + \alpha/\Gamma_{imp}}{ (1 + \alpha/\Gamma_{imp})^{2} + (\omega_{c}/\Gamma_{imp})^{2}} = \sigma_{imp} \frac{1 \pm \frac{1}{\pi N_{F}} \int \frac{d^{3} \bm{q}}{(2\pi)^{3}} \frac{1}{D \bm{q}^{2} + \tau_{\phi}^{-1}}}{ \Bigl(1 \pm \frac{1}{\pi N_{F}} \int \frac{d^{3} \bm{q}}{(2\pi)^{3}} \frac{1}{D \bm{q}^{2} + \tau_{\phi}^{-1}} \Bigr)^{2} + (\omega_{c}/\Gamma_{imp})^{2}} \eqa and
\bqa && \sigma_{yx} = - \sigma_{imp} \frac{\omega_{c}/\Gamma_{imp}}{(1 + \alpha/\Gamma_{imp})^{2} + (\omega_{c}/\Gamma_{imp})^{2}} = - \sigma_{imp} \frac{\omega_{c} / \Gamma_{imp}}{\Bigl(1 \pm \frac{1}{\pi N_{F}} \int \frac{d^{3} \bm{q}}{(2\pi)^{3}} \frac{1}{D \bm{q}^{2} + \tau_{\phi}^{-1}}\Bigr)^{2} + \omega_{c}^{2} / \Gamma_{imp}^{2}} , \eqa quite familiar except for the weak-localization (weak anti-localization) correction. Inverting the denominator with the numerator in Eq. (25) and resorting to the Einstein relation $\sigma_{imp} = 2 e^{2} N_{F} D_{imp}$, we recover the well-known weak-localization (weak anti-localization) formula \cite{Review_Disorder} for the magneto-resistivity ($\rho_{xx} \approx \frac{1}{\sigma_{xx}}$ ) \bqa && \rho_{xx} = \rho_{imp} \pm \mathcal{C} e^{2} N_{F} \rho_{imp}^{2} \int^{1/l_{imp}}_{1/l_{ph}} d q q^{2} \frac{1}{ \bm{q}^{2}} , \eqa where $\pm$ corresponds to weak (anti-)localization and the part of the cyclotron frequency is neglected in the weak-field limit. $\rho_{imp} = \frac{1}{\sigma_{imp}}$ is a residual resistivity due to elastic impurity scattering and $\mathcal{C}$ is a positive numerical constant. $l_{imp}$ in the upper cut-off is the mean-free path and $l_{ph}$ in the lower cutoff is the phase-coherent length, as discussed before. If one sets $l_{ph}^{-1} \propto \sqrt{B}$ in the lower cut-off, we reproduce the magneto-resistivity with weak (anti-)localization \cite{Review_Disorder}.

An interesting result is that the Hall conductivity encodes the weak-localization (anti-localization) quantum correction, not discussed before as far as we know. This correction gives rise to an unexpected behavior for the Hall conductivity. For example, we find that it vanishes with a logarithmic correction in two dimensions as we approach zero magnetic field, given by \bqa && \sigma_{yx}(B) \propto B [\ln (B/B_{0})]^{-2} , \eqa where $B_{0}$ is a scale of magnetic field, coming from the upper cut-off. In three dimensions, we may observe deviation from the linear dependence of magnetic field, expected to cause confusion with an anomalous Hall signal. In spite of this quantum correction, the Hall resistivity recovers the well-known formula, given by $\rho_{yx} = \sigma_{yx} / (\sigma_{xx}^{2} + \sigma_{yx}^{2}) = - 1/ (n e c)$, which seems to justify our derivation. We believe that this subject needs to be investigated more sincerely for various samples showing weak-localization (weak anti-localization) corrections.

It is straightforward to obtain the optical magneto-conductivity and the optical Hall coefficient with the weak-localization (anti-localization) quantum correction. Although we do not discuss these aspects more, it will be interesting to observe the regime that shows such quantum corrections clearly in optical responses.

\subsection{Formulation}

Introducing both weak anti-localization quantum corrections through the collision term and topological structures through the semi-classical equation of motion into the Boltzmann-equation framework, we reach our starting point for anomalous transport phenomena in Weyl metal, where an effective theory is given by
\bqa && \Bigl\{ - i \nu + \Bigl( 1 + \frac{e}{c} \boldsymbol{B} \cdot \boldsymbol{\Omega}_{\boldsymbol{p}}^{\chi} \Bigr)^{-1} \Bigl( e \boldsymbol{E} + \frac{e}{c} \boldsymbol{v}_{\boldsymbol{p}} \times \boldsymbol{B}
+ \frac{e^{2}}{c} (\boldsymbol{E} \cdot \boldsymbol{B}) \boldsymbol{\Omega}_{\boldsymbol{p}}^{\chi} \Bigr) \cdot \bm{\nabla}_{\bm{p}} \Bigr\} f_{\chi}(\bm{p};\nu) \nn && = - \Gamma_{imp} [f_{\chi}(\bm{p};\nu) - f_{eq}(\bm{p})] - \Gamma'_{imp} [f_{\chi}(\bm{p};\nu) - f_{-\chi}(\bm{p};\nu)] - \alpha_{\chi}(\nu) [f_{\chi}(-\bm{p};\nu) - f_{eq}(\bm{p})] , \eqa where $\chi = \pm$ represents each chirality. An important point, not discussed explicitly in the introduction, is to introduce an inter Weyl-point scattering term into the Boltzmann equation phenomenologically, where the relaxation rate for the inter-node scattering is $\Gamma_{imp}'$. The weak anti-localization kernel is given by
\bqa && \alpha_{\chi}(\nu) = - \frac{\Gamma_{imp}+\Gamma'_{imp}}{\pi N_{F}} \int \frac{d^{3} \bm{q}}{(2\pi)^{3}} \frac{1}{D_{\chi} \bm{q}^{2} - i \nu + \tau_{\phi}^{-1}} , \eqa where $D_{\chi}$ is the diffusion coefficient for each Weyl point, assumed to be identical, i.e., $D_{+} = D_{-} = D$.

Solving these coupled Boltzmann equations, we obtain the expression for an electric current, given by
\bqa && \bm{j}(\nu) = e \int \frac{d^{3} \bm{p}}{(2\pi)^{3}} \Bigl\{ \Bigl( 1 + \frac{e}{c}
\boldsymbol{B} \cdot \boldsymbol{\Omega}_{\boldsymbol{p}}^{+} \Bigr) \dot{\bm r}_{+} f_{+}(\bm{p};\nu) + \Bigl( 1 + \frac{e}{c}
\boldsymbol{B} \cdot \boldsymbol{\Omega}_{\boldsymbol{p}}^{-} \Bigr) \dot{\bm r}_{-} f_{-}(\bm{p};\nu) \Bigr\} \nn && = e \int \frac{d^{3} \bm{p}}{(2\pi)^{3}} \Bigl\{ \boldsymbol{v}_{\boldsymbol{p}} + e \boldsymbol{E} \times \boldsymbol{\Omega}_{\boldsymbol{p}}^{+} + \frac{e}{c} \boldsymbol{\Omega}_{\boldsymbol{p}}^{+} \cdot
\boldsymbol{v}_{\boldsymbol{p}} \boldsymbol{B} \Bigr\} f_{+}(\bm{p};\nu) + e \int \frac{d^{3} \bm{p}}{(2\pi)^{3}} \Bigl\{ \boldsymbol{v}_{\boldsymbol{p}} + e \boldsymbol{E} \times \boldsymbol{\Omega}_{\boldsymbol{p}}^{-} + \frac{e}{c} \boldsymbol{\Omega}_{\boldsymbol{p}}^{-} \cdot \boldsymbol{v}_{\boldsymbol{p}} \boldsymbol{B} \Bigr\} f_{-}(\bm{p};\nu) . \eqa

\subsection{$\bm{E} = E \hat{\bm x}$ and $\bm{B} = B \hat{\bm z}$}

In order to clarify the role of the ``topological" $\bm{E} \cdot \bm{B}$ term, it is necessary to evaluate transport coefficients in the normal setup of $\bm{E} = E \hat{\bm x}$ and $\bm{B} = B \hat{\bm z}$. Here, `` " is utilized to mean that this $\bm{E} \cdot \bm{B}$ term is not topological any more since it is introduced in the equation of motion, originating from the space-time dependence of the $\theta(\bm{r},t)$ coefficient, where the origin of this term is topological.

We start from the following coupled Boltzmann equations in the linear-response regime and the dc-limit,
\bqa && \Bigl\{\Gamma_{imp} + \Gamma_{imp}' + \alpha - \Bigl( 1 + \frac{e B}{c} \Omega_{z}^{\chi}(\bm{p}) \Bigr)^{-1} \frac{e B}{c} \Bigl( v_{x}(\bm{p}) \frac{\partial}{\partial p_{y}} - v_{y}(\bm{p}) \frac{\partial}{\partial p_{x}} \Bigr)  \Bigr\} f_{\chi}(\bm{p}) \nn && = [\Gamma_{imp} + \Gamma_{imp}' + \alpha] f_{eq}(\bm{p}) - \Bigl( 1 + \frac{e B}{c} \Omega_{z}^{\chi}(\bm{p}) \Bigr)^{-1} e E \frac{\partial }{\partial p_{x}} f_{eq} (\bm{p}) + \Gamma'_{imp}[f_{-\chi}(\bm{p}) - f_{eq} (\bm{p})] , \eqa where the $\bm{E} \cdot \bm{B}$ term disappears. These equations lead us to consider the ansatz below
\bqa && f_{\chi}(\bm{p}) = f_{eq}(\bm{p}) - \frac{\Bigl( 1 + \frac{e B}{c} \Omega_{z}^{\chi}(\bm{p}) \Bigr)^{-1}}{\Gamma_{imp} + \Gamma'_{imp} + \alpha} e E \frac{\partial }{\partial p_{x}} f_{eq} (\bm{p}) + \Bigl( - \frac{\partial}{\partial \epsilon} f_{eq} (\epsilon) \Bigr) \bm{v}_{\bm{p}} \cdot \bm{\Lambda}_{\chi}(\bm{p}) . \eqa

It is natural to assume \bqa && |\Gamma'_{imp}| \ll \Gamma_{imp} , \eqa where the distance between a paired Weyl points gives rise to a smaller relaxation rate for the inter-node scattering than that for the intra-node one in the case of charged impurities. However, it is straightforward to consider $\delta^{(3)}(\bm{r})$-type potentials in this Boltzmann-equation framework. In this paper we focus on the case of charged impurities for simplicity. Then, these coupled Boltzmann equations become simplified as follows
%
%\bqa &&  - \Bigl( 1 + \frac{e B}{c} \Omega_{z}^{+}(\bm{p}) \Bigr)^{-1} \frac{e B}{m c} \Bigl( v_{x}(\bm{p}) \Lambda_{y}^{+}(\bm{p}) - v_{y}(\bm{p}) \Lambda_{x}^{+}(\bm{p}) \Bigr) + [\Gamma_{imp} + \alpha] \bm{v}_{\bm{p}} \cdot \bm{\Lambda}_{+}(\bm{p}) \nn && + \Gamma_{imp}' \Bigl\{ - \frac{\Bigl( 1 + \frac{e B}{c} \Omega_{z}^{-}(\bm{p}) \Bigr)^{-1}}{\Gamma_{imp} + \alpha} e E \frac{\partial }{\partial p_{x}} f_{eq} (\bm{p}) + \Bigl( - \frac{\partial}{\partial \epsilon} f_{eq} (\epsilon) \Bigr) \bm{v}_{\bm{p}} \cdot \bm{\Lambda}_{-}(\bm{p}) \Bigr\} = 0 , \nn && - \Bigl( 1 + \frac{e B}{c} \Omega_{z}^{-}(\bm{p}) \Bigr)^{-1} \frac{e B}{m c} \Bigl( v_{x}(\bm{p}) \Lambda_{y}^{-}(\bm{p}) - v_{y}(\bm{p}) \Lambda_{x}^{-}(\bm{p}) \Bigr) + [\Gamma_{imp} + \alpha] \bm{v}_{\bm{p}} \cdot \bm{\Lambda}_{-}(\bm{p}) \nn && + \Gamma_{imp}' \Bigl\{ - \frac{\Bigl( 1 + \frac{e B}{c} \Omega_{z}^{+}(\bm{p}) \Bigr)^{-1}}{\Gamma_{imp} + \alpha} e E \frac{\partial }{\partial p_{x}} f_{eq} (\bm{p}) + \Bigl( - \frac{\partial}{\partial \epsilon} f_{eq} (\epsilon) \Bigr) \bm{v}_{\bm{p}} \cdot \bm{\Lambda}_{+}(\bm{p}) \Bigr\} = 0 . \eqa
%
\bqa &&  - \Bigl( 1 + \frac{e B}{c} \Omega_{z}^{\chi}(\bm{p}) \Bigr)^{-1} \frac{e B}{m c} \Bigl( v_{x}(\bm{p}) \Lambda_{y}^{\chi}(\bm{p}) - v_{y}(\bm{p}) \Lambda_{x}^{\chi}(\bm{p}) \Bigr) + [\Gamma_{imp} + \alpha] \bm{v}_{\bm{p}} \cdot \bm{\Lambda}_{\chi}(\bm{p}) \nn && + \Bigl( 1 + \frac{e B}{c} \Omega_{z}^{\chi}(\bm{p}) \Bigr)^{-2} \frac{e B}{m c} \frac{e E v_{y}(\bm{p})}{\Gamma_{imp} + \alpha} - \frac{\Gamma_{imp}'}{\Gamma_{imp} + \alpha} \Bigl( 1 + \frac{e B}{c} \Omega_{z}^{-\chi}(\bm{p}) \Bigr)^{-1} e E v_{x}(\bm{p}) \approx 0 , \eqa where only the linear order in $\Gamma_{imp}' / \Gamma_{imp}$ is kept, allowing us to decouple these equations.

The solution of $\bm{\Lambda}_{\chi}(\bm{p})$ is determined from the condition that these Boltzmann equations must be satisfied for any values of velocity. It is convenient to rewrite such Boltzmann equations as follows \bqa && \Re \Bigl\{ \Bigl( \Gamma_{imp} - i \Omega_{c}^{\chi}(\bm{p}) + \alpha \Bigr) V(\bm{p}) \Lambda_{\chi}(\bm{p}) - i \Bigl( 1 + \frac{e B}{c} \Omega_{z}^{\chi}(\bm{p}) \Bigr)^{-1} \frac{e E \Omega_{c}^{\chi}(\bm{p})}{\Gamma_{imp} + \alpha} V(\bm{p}) - \frac{\Gamma_{imp}'}{\Gamma_{imp} + \alpha} \Bigl( 1 + \frac{e B}{c} \Omega_{z}^{-\chi}(\bm{p}) \Bigr)^{-1} e E V(\bm{p}) \Bigr\} = 0 , \nn \eqa introducing $V(\bm{p}) = v_{x}(\bm{p}) + i v_{y}(\bm{p})$ and $\Lambda_{\chi}(\bm{p}) = \Lambda_{x}^{\chi}(\bm{p}) - i \Lambda_{y}^{\chi}(\bm{p})$ into them, where $\Omega_{c}^{\chi}(\bm{p}) = \Bigl( 1 + \frac{e B}{c} \Omega_{z}^{\chi}(\bm{p}) \Bigr)^{-1} \frac{e B}{m c}$ is an effective cyclotron frequency around each Weyl point. Then, we find
\bqa && \Lambda_{x}^{\chi}(\bm{p}) = - e E \frac{1}{\Gamma_{imp} + \alpha} \frac{\Bigl( 1 + \frac{e B}{c} \Omega_{z}^{\chi}(\bm{p}) \Bigr)^{-1} [\Omega_{c}^{\chi}(\bm{p})]^{2} - \Bigl( 1 + \frac{e B}{c} \Omega_{z}^{-\chi}(\bm{p}) \Bigr)^{-1} \Gamma_{imp}' [\Gamma_{imp} + \alpha]}{ [\Gamma_{imp} + \alpha]^{2} + [\Omega_{c}^{\chi}(\bm{p})]^{2}} \eqa and
\bqa && \Lambda_{y}^{\chi}(\bm{p}) = - e E \frac{1}{\Gamma_{imp} + \alpha} \frac{\Bigl( 1 + \frac{e B}{c} \Omega_{z}^{\chi}(\bm{p}) \Bigr)^{-1} \Omega_{c}^{\chi}(\bm{p}) [\Gamma_{imp} + \alpha] + \Bigl( 1 + \frac{e B}{c} \Omega_{z}^{-\chi}(\bm{p}) \Bigr)^{-1} \Omega_{c}^{\chi}(\bm{p}) \Gamma_{imp}'}{[\Gamma_{imp} + \alpha]^{2} + [ \Omega_{c}^{\chi}(\bm{p})]^{2}} . \eqa
As a result, each distribution function is given by
\bqa && f_{\chi}(\bm{p}) = f_{eq}(\bm{p}) + \Bigl( - \frac{\partial }{\partial \epsilon} f_{eq} (\epsilon) \Bigr) \Bigl( 1 + \frac{e B}{c} \Omega_{z}^{\chi}(\bm{p}) \Bigr)^{-1} \frac{1}{\Gamma_{imp} + \alpha} e E v_{x}(\bm{p}) \nn && - \Bigl( - \frac{\partial}{\partial \epsilon} f_{eq} (\epsilon) \Bigr) \frac{1}{\Gamma_{imp} + \alpha} \frac{\Bigl( 1 + \frac{e B}{c} \Omega_{z}^{\chi}(\bm{p}) \Bigr)^{-1} [\Omega_{c}^{\chi}(\bm{p})]^{2} - \Bigl( 1 + \frac{e B}{c} \Omega_{z}^{-\chi}(\bm{p}) \Bigr)^{-1} \Gamma_{imp}' [\Gamma_{imp} + \alpha]}{ [\Gamma_{imp} + \alpha]^{2} + [\Omega_{c}^{\chi}(\bm{p})]^{2}} e E v_{x}({\bm{p}}) \nn && - \Bigl( - \frac{\partial}{\partial \epsilon} f_{eq} (\epsilon) \Bigr) \frac{1}{\Gamma_{imp} + \alpha} \frac{\Bigl( 1 + \frac{e B}{c} \Omega_{z}^{\chi}(\bm{p}) \Bigr)^{-1} \Omega_{c}^{\chi}(\bm{p}) [\Gamma_{imp} + \alpha] + \Bigl( 1 + \frac{e B}{c} \Omega_{z}^{-\chi}(\bm{p}) \Bigr)^{-1} \Omega_{c}^{\chi}(\bm{p}) \Gamma_{imp}'}{[\Gamma_{imp} + \alpha]^{2} + [ \Omega_{c}^{\chi}(\bm{p})]^{2}} e E v_{y}({\bm{p}}) . \eqa

Inserting these formulae into the current formulae, we obtain the magneto-conductivity
\bqa && \sigma_{xx}^{\chi} = e^{2} \int \frac{d^{3} \bm{p}}{(2\pi)^{3}} [v_{x}(\bm{p})]^{2} \Bigl( - \frac{\partial}{\partial \epsilon} f_{eq} (\epsilon) \Bigr) \Bigl( 1 + \frac{e B}{c} \Omega_{z}^{\chi}(\bm{p}) \Bigr)^{-1} \Bigl\{ \frac{1}{\Gamma_{imp} + \alpha} - \frac{1}{\Gamma_{imp} + \alpha} \frac{[\Omega_{c}^{\chi}(\bm{p})]^{2}}{ [\Gamma_{imp} + \alpha]^{2} + [\Omega_{c}^{\chi}(\bm{p})]^{2}} \Bigr\} \nn && + e^{2} \int \frac{d^{3} \bm{p}}{(2\pi)^{3}} [v_{x}(\bm{p})]^{2} \Bigl( - \frac{\partial}{\partial \epsilon} f_{eq} (\epsilon) \Bigr) \Bigl( 1 + \frac{e B}{c} \Omega_{z}^{-\chi}(\bm{p}) \Bigr)^{-1} \frac{\Gamma_{imp}'}{[\Gamma_{imp} + \alpha]^{2} + [\Omega_{c}^{\chi}(\bm{p})]^{2}} \eqa and the Hall conductivity
\bqa && \sigma_{yx}^{\chi} = - e^{2} \int \frac{d^{3} \bm{p}}{(2\pi)^{3}} \Omega_{z}^{\chi}(\bm{p}) f_{eq}(\bm{p}) \nn && - e^{2} \int \frac{d^{3} \bm{p}}{(2\pi)^{3}} [v_{y}(\bm{p})]^{2} \Bigl( - \frac{\partial}{\partial \epsilon} f_{eq} (\epsilon) \Bigr) \Bigl( 1 + \frac{e B}{c} \Omega_{z}^{\chi}(\bm{p}) \Bigr)^{-1} \frac{\Omega_{c}^{\chi}(\bm{p})}{[\Gamma_{imp} + \alpha]^{2}+ [\Omega_{c}^{\chi}(\bm{p})]^{2}} \nn && - e^{2} \int \frac{d^{3} \bm{p}}{(2\pi)^{3}} [v_{y}(\bm{p})]^{2} \Bigl( - \frac{\partial}{\partial \epsilon} f_{eq} (\epsilon) \Bigr) \Bigl( 1 + \frac{e B}{c} \Omega_{z}^{-\chi}(\bm{p}) \Bigr)^{-1} \frac{\Gamma_{imp}'}{\Gamma_{imp} + \alpha} \frac{ \Omega_{c}^{\chi}(\bm{p})}{[\Gamma_{imp} + \alpha]^{2} + [ \Omega_{c}^{\chi}(\bm{p})]^{2}} \eqa around each Weyl point.

The momentum integral can be performed in a formal way, resorting to $\bm{\Omega}_{\bm{p}}^{+} \approx - \bm{\Omega}_{\bm{p}}^{-} = \bm{\Omega}_{\bm{p}}$, which gives rise to cancelation for linear terms in Berry curvature. Then, the magneto-conductivity is given by
\bqa && \sigma_{xx} \approx 2 \sigma \frac{1 + \alpha/\Gamma_{imp} + \Gamma_{imp}' / \Gamma_{imp}}{ [1 + \alpha/\Gamma_{imp}]^{2} + \omega_{c}^{2}/\Gamma_{imp}^{2}} , \eqa where the Drude conductivity $\sigma$ is defined in a similar way of the previous section while $2$ comes from two Weyl cones. This expression reads \bqa && \rho_{xx} \approx \Bigl( 1 - \frac{\Gamma_{imp}'}{\Gamma_{imp}} \Bigr) \rho_{imp} - \mathcal{C} e^{2} N_{F} \rho_{imp}^{2} \int^{1/l_{imp}}_{1/l_{ph}} d q q^{2} \frac{1}{ \bm{q}^{2}} \eqa in the leading order for magnetic field, where $l_{ph}^{-1} \propto \sqrt{B}$ as discussed before.
%Of course, the inter-node scattering enhances the residual electrical resistivity. The weak anti-localization correction remains unaltered.

The Hall conductivity is \bqa && \sigma_{yx} = - e^{2} \int \frac{d^{3} \bm{p}}{(2\pi)^{3}} [\Omega_{z}^{+}(\bm{p}) + \Omega_{z}^{-}(\bm{p})] f_{eq}(\bm{p}) - 2 \sigma \frac{ \omega_{c} / \Gamma_{imp} }{ (1 + \alpha/\Gamma_{imp})^{2} + (\omega_{c}/\Gamma_{imp})^{2} } \frac{1 + \alpha/\Gamma_{imp} + \Gamma_{imp}' / \Gamma_{imp}}{1 + \alpha/\Gamma_{imp}} , \eqa where the first term is an anomalous contribution resulting from the Berry curvature \cite{Semiclassical_Eqs1,Semiclassical_Eqs2}. Inserting $\bm{\Omega}_{\bm{p}}^{\chi} \propto \chi \frac{\bm{\hat{p}}}{|\bm{p} - \chi g_{\psi} \bm{B}|^{2}}$ with $\bm{B} = B \bm{\hat{z}}$ and $\chi = \pm$ into the expression of the anomalous Hall coefficient and performing the momentum integration, we find that it is proportional to the momentum-space distance between the pair of Weyl points, i.e., $g_{\psi} B$, consistent with that based on the diagrammatic analysis \cite{AHE_Kubo1,AHE_Kubo2}. For the normal contribution, the presence of the inter-node scattering modifies the Hall coefficient as follows
\bqa && \rho_{yx} = \frac{\sigma_{yx}}{\sigma_{xx}^{2} + \sigma_{yx}^{2}} = - \frac{1}{n e c} \Bigl(1 + \frac{\Gamma_{imp}'}{\Gamma_{imp}} \frac{1}{1 + \alpha/\Gamma_{imp}} \Bigr) , \eqa which turns out to be not a constant but a function of magnetic field, combined with the weak anti-localization correction.

\subsection{$\bm{E} = E \hat{\bm x}$ and $\bm{B} = B \hat{\bm x}$}

Our main problem is to investigate both the magneto-conductivity and Hall conductivity when electric field is applied in parallel with magnetic field, i.e., the case of $\bm{E} = E \hat{\bm x}$ and $\bm{B} = B \hat{\bm x}$. Coupled Boltzmann equations are given by
\bqa && \Bigl\{\Gamma_{imp} + \Gamma_{imp}' + \alpha - \Bigl( 1 + \frac{e B}{c} \Omega_{x}^{\chi}(\bm{p}) \Bigr)^{-1} \frac{e B}{c} \Bigl( v_{y}(\bm{p}) \frac{\partial}{\partial p_{z}} - v_{z}(\bm{p}) \frac{\partial}{\partial p_{y}} \Bigr) \Bigr\} f_{\chi}(\bm{p}) \nn && = [\Gamma_{imp} + \Gamma_{imp}' + \alpha] f_{eq}(\bm{p}) - \Bigl( 1 + \frac{e B}{c} \Omega_{x}^{\chi}(\bm{p}) \Bigr)^{-1} \Bigl( e E \frac{\partial }{\partial p_{x}} f_{eq} (\bm{p}) + \frac{e^{2}}{c} (E B) \boldsymbol{\Omega}_{\boldsymbol{p}}^{\chi} \cdot \bm{\nabla}_{\bm{p}} f_{eq} (\bm{p}) \Bigr) + \Gamma'_{imp}[f_{-\chi}(\bm{p}) - f_{eq} (\bm{p})] . \nn \eqa An essential aspect is the existence of the $\bm{E} \cdot \bm{B}$ term, which plays the role of an additional force beyond the Lorentz force, giving rise to not only an additional drift along the direction of electric field but also a transverse motion along the $\bm{y}$-direction associated with the direction of Berry curvature. The former results in negative magneto-resistivity while the latter causes an anomalous Hall effect that has nothing to do with the ``conventional" anomalous Hall effect \cite{Semiclassical_Eqs1,Semiclassical_Eqs2} in the previous section. However, this novel anomalous Hall effect turns out to be canceled when each Weyl-point contribution is summed.

Following the previous strategy, we take the ansatz
\bqa && f_{\chi}(\bm{p}) = f_{eq}(\bm{p}) - \frac{\Bigl( 1 + \frac{e B}{c} \Omega_{x}^{\chi}(\bm{p}) \Bigr)^{-1}}{\Gamma_{imp} + \Gamma'_{imp} + \alpha} \Bigl( e E \frac{\partial }{\partial p_{x}} f_{eq} (\bm{p}) + \frac{e^{2}}{c} (E B) \boldsymbol{\Omega}_{\boldsymbol{p}}^{\chi} \cdot \bm{\nabla}_{\bm{p}} f_{eq} (\bm{p}) \Bigr) + \Bigl( - \frac{\partial}{\partial \epsilon} f_{eq} (\epsilon) \Bigr) \bm{v}_{\bm{p}} \cdot \bm{\Lambda}_{\chi}(\bm{p}) , \eqa where the $\bm{E} \cdot \bm{B}$ term exists.

Resorting to $\Gamma'_{imp} \ll \Gamma_{imp}$ and keeping the linear order for $\Gamma_{imp}'$, we obtain
\bqa && - \frac{\Bigl( 1 + \frac{e B}{c} \Omega_{x}^{\chi}(\bm{p}) \Bigr)^{-2}}{\Gamma_{imp} + \alpha} \frac{e^{2}}{c} (E B) \frac{e B}{m c} \Bigl( \Omega_{z}^{\chi}(\bm{p}) v_{y}(\bm{p}) - \Omega_{y}^{\chi}(\bm{p}) v_{z}(\bm{p}) \Bigr) \nn && - \Bigl( 1 + \frac{e B}{c} \Omega_{x}^{\chi}(\bm{p}) \Bigr)^{-1} \frac{e B}{m c} \Bigl( v_{y}(\bm{p}) \Lambda_{z}^{\chi}(\bm{p}) - v_{z}(\bm{p}) \Lambda_{y}^{\chi}(\bm{p}) \Bigr) + [\Gamma_{imp} + \alpha] \bm{v}_{\bm{p}} \cdot \bm{\Lambda}_{\chi}(\bm{p})
\nn && - \frac{\Gamma_{imp}'}{\Gamma_{imp} + \alpha} \Bigl( 1 + \frac{e B}{c} \Omega_{x}^{-\chi}(\bm{p}) \Bigr)^{-1} \Bigl( e E v_{x}(\bm{p}) + \frac{e^{2}}{c} (E B) \boldsymbol{\Omega}_{\boldsymbol{p}}^{-\chi} \cdot \bm{v}_{\bm{p}} \Bigr) \approx 0 , \eqa which allows us to decouple the Boltzmann equations for $\bm{\Lambda}_{\pm}(\bm{p})$.

It is easy to find $\Lambda_{x}^{\chi}(\bm{p})$ since they are not coupled with $\Lambda_{y,z}^{\chi}(\bm{p})$, given by
\bqa && \Lambda_{x}^{\chi}(\bm{p}) = \frac{\Gamma_{imp}' }{ [\Gamma_{imp} + \alpha]^{2}} \Bigl( 1 + \frac{e B}{c} \Omega_{x}^{-\chi}(\bm{p}) \Bigr)^{-1} \Bigl( e E + \frac{e^{2}}{c} (E B) \Omega_{x}^{-\chi}(\bm{p}) \Bigr) . \eqa On the other hand, $\Lambda_{y}^{\chi}(\bm{p})$ are coupled with $\Lambda_{z}^{\chi}(\bm{p})$, giving rise to complications. Introducing complex notations \bqa && V(\bm{p}) = v_{y}(\bm{p}) + i v_{z}(\bm{p}) , ~~~~~ \Lambda_{\chi}(\bm{p}) = \Lambda_{y}^{\chi}(\bm{p}) - i \Lambda_{z}^{\chi}(\bm{p}) , ~~~~~ \Omega_{\chi}(\bm{p}) = \Omega_{y}^{\chi}(\bm{p}) - i \Omega_{z}^{\chi}(\bm{p}) , \eqa we rewrite the above expression as follows
\bqa && \Re \Bigl\{ \Bigl( \Gamma_{imp} - i \Omega_{c}^{\chi}(\bm{p}) + \alpha \Bigr) V(\bm{p}) \Lambda_{\chi}(\bm{p}) - i \Bigl( 1 + \frac{e B}{c} \Omega_{x}^{\chi}(\bm{p}) \Bigr)^{-1} \frac{\frac{e^{2}}{c} (E B) \Omega_{c}^{\chi}(\bm{p})}{\Gamma_{imp} + \alpha} V(\bm{p}) \Omega_{\chi}(\bm{p}) \nn && - \frac{\Gamma_{imp}'}{\Gamma_{imp} + \alpha} \Bigl( 1 + \frac{e B}{c} \Omega_{x}^{-\chi}(\bm{p}) \Bigr)^{-1} \frac{e^{2}}{c} (E B) V(\bm{p}) \Omega_{-\chi}(\bm{p}) \Bigr\} = 0 , \eqa where $\Omega_{c}^{\chi}(\bm{p}) = \Bigl( 1 + \frac{e B}{c} \Omega_{x}^{\chi}(\bm{p}) \Bigr)^{-1} \frac{e B}{m c}$ is an effective cyclotron frequency. Actually, the structure of this equation is quite similar to that of the previous section, where $e E$ is replaced with $\frac{e^{2}}{c} (E B)$ with the Berry curvature $\Omega_{\chi}(\bm{p})$. It is straightforward to find the solution, given by
\bqa && \Lambda_{y}^{\chi}(\bm{p}) = - \frac{e^{2}}{c} (E B) \Bigl( 1 + \frac{e B}{c} \Omega_{x}^{\chi}(\bm{p}) \Bigr)^{-1} \frac{1}{\Gamma_{imp} + \alpha} \frac{ [\Omega_{c}^{\chi}(\bm{p})]^{2} \Omega_{y}^{\chi}(\bm{p}) - (\Gamma_{imp} + \alpha) \Omega_{c}^{\chi}(\bm{p}) \Omega_{z}^{\chi}(\bm{p})}{ [\Gamma_{imp} + \alpha]^{2} + [ \Omega_{c}^{\chi}(\bm{p})]^{2} } \nn && + \frac{e^{2}}{c} (E B) \Bigl( 1 + \frac{e B}{c} \Omega_{x}^{-\chi}(\bm{p}) \Bigr)^{-1} \frac{\Gamma_{imp}'}{\Gamma_{imp} + \alpha} \frac{(\Gamma_{imp} + \alpha) \Omega_{y}^{-\chi}(\bm{p}) + \Omega_{c}^{\chi}(\bm{p}) \Omega_{z}^{-\chi}(\bm{p})}{ [\Gamma_{imp} + \alpha]^{2} + [ \Omega_{c}^{\chi}(\bm{p})]^{2} }  \eqa
and \bqa && \Lambda_{z}^{\chi}(\bm{p}) = - \frac{e^{2}}{c} (E B) \Bigl( 1 + \frac{e B}{c} \Omega_{x}^{\chi}(\bm{p}) \Bigr)^{-1} \frac{1}{\Gamma_{imp} + \alpha} \frac{ (\Gamma_{imp} + \alpha) \Omega_{c}^{\chi}(\bm{p}) \Omega_{y}^{\chi}(\bm{p}) + [\Omega_{c}^{\chi}(\bm{p})]^{2} \Omega_{z}^{\chi}(\bm{p})}{ [\Gamma_{imp} + \alpha]^{2} + [ \Omega_{c}^{\chi}(\bm{p})]^{2} } \nn && - \frac{e^{2}}{c} (E B) \Bigl( 1 + \frac{e B}{c} \Omega_{x}^{-\chi}(\bm{p}) \Bigr)^{-1} \frac{\Gamma_{imp}'}{\Gamma_{imp} + \alpha} \frac{\Omega_{c}^{\chi}(\bm{p}) \Omega_{y}^{-\chi}(\bm{p}) - (\Gamma_{imp} + \alpha) \Omega_{z}^{-\chi}(\bm{p})}{ [\Gamma_{imp} + \alpha]^{2} + [ \Omega_{c}^{\chi}(\bm{p})]^{2} } . \eqa An interesting point is that these corrections are proportional to $\bm{E}\cdot\bm{B}$. As discussed before, such an $\bm{E}\cdot\bm{B}$ term gives rise to an additional force-like term besides the Lorentz force.

Inserting these corrections into the ansatz of the distribution function, we obtain
\bqa && f_{\chi}(\bm{p}) = f_{eq}(\bm{p}) + \Bigl( - \frac{\partial }{\partial \epsilon} f_{eq} (\epsilon) \Bigr) \Bigl( 1 + \frac{e B}{c} \Omega_{x}^{\chi}(\bm{p}) \Bigr)^{-1} \frac{1}{\Gamma_{imp} + \alpha} \Bigl( e E v_{x}(\bm{p}) + \frac{e^{2}}{c} (E B) \boldsymbol{\Omega}_{\boldsymbol{p}}^{\chi} \cdot \bm{v}_{\bm{p}} \Bigr) \nn && + \Bigl( - \frac{\partial}{\partial \epsilon} f_{eq} (\epsilon) \Bigr) \Bigl( 1 + \frac{e B}{c} \Omega_{x}^{-\chi}(\bm{p}) \Bigr)^{-1} \frac{\Gamma_{imp}' }{[\Gamma_{imp} + \alpha]^{2}} \Bigl( e E v_{x}(\bm{p}) + \frac{e^{2}}{c} (E B) \Omega_{x}^{-\chi}(\bm{p}) v_{x}(\bm{p}) \Bigr) \nn && - \Bigl( - \frac{\partial}{\partial \epsilon} f_{eq} (\epsilon) \Bigr) \Bigl( 1 + \frac{e B}{c} \Omega_{x}^{\chi}(\bm{p}) \Bigr)^{-1} v_{y}({\bm{p}}) \frac{e^{2}}{c} (E B) \frac{1}{\Gamma_{imp} + \alpha} \frac{ [\Omega_{c}^{\chi}(\bm{p})]^{2} \Omega_{y}^{\chi}(\bm{p}) - (\Gamma_{imp} + \alpha) \Omega_{c}^{\chi}(\bm{p}) \Omega_{z}^{\chi}(\bm{p})}{ [\Gamma_{imp} + \alpha]^{2} + [ \Omega_{c}^{\chi}(\bm{p})]^{2} } \nn && + \Bigl( - \frac{\partial}{\partial \epsilon} f_{eq} (\epsilon) \Bigr) \Bigl( 1 + \frac{e B}{c} \Omega_{x}^{-\chi}(\bm{p}) \Bigr)^{-1} v_{y}({\bm{p}}) \frac{e^{2}}{c} (E B) \frac{\Gamma_{imp}'}{\Gamma_{imp} + \alpha} \frac{(\Gamma_{imp} + \alpha) \Omega_{y}^{-\chi}(\bm{p}) + \Omega_{c}^{\chi}(\bm{p}) \Omega_{z}^{-\chi}(\bm{p})}{ [\Gamma_{imp} + \alpha]^{2} + [ \Omega_{c}^{\chi}(\bm{p})]^{2} } \nn && - \Bigl( - \frac{\partial}{\partial \epsilon} f_{eq} (\epsilon) \Bigr) \Bigl( 1 + \frac{e B}{c} \Omega_{x}^{\chi}(\bm{p}) \Bigr)^{-1} v_{z}({\bm{p}}) \frac{e^{2}}{c} (E B) \frac{1}{\Gamma_{imp} + \alpha} \frac{ (\Gamma_{imp} + \alpha) \Omega_{c}^{\chi}(\bm{p}) \Omega_{y}^{\chi}(\bm{p}) + [\Omega_{c}^{\chi}(\bm{p})]^{2} \Omega_{z}^{\chi}(\bm{p})}{ [\Gamma_{imp} + \alpha]^{2} + [ \Omega_{c}^{\chi}(\bm{p})]^{2} } \nn && - \Bigl( - \frac{\partial}{\partial \epsilon} f_{eq} (\epsilon) \Bigr) \Bigl( 1 + \frac{e B}{c} \Omega_{x}^{-\chi}(\bm{p}) \Bigr)^{-1} v_{z}({\bm{p}}) \frac{e^{2}}{c} (E B) \frac{\Gamma_{imp}'}{\Gamma_{imp} + \alpha} \frac{\Omega_{c}^{\chi}(\bm{p}) \Omega_{y}^{-\chi}(\bm{p}) - (\Gamma_{imp} + \alpha) \Omega_{z}^{-\chi}(\bm{p})}{ [\Gamma_{imp} + \alpha]^{2} + [ \Omega_{c}^{\chi}(\bm{p})]^{2} } . \eqa Although these expressions look complicated, an essential modification compared with those of the previous normal setup lies in the $\bm{E}\cdot\bm{B}$ term. In particular, the contribution of the $\bm{E}\cdot\bm{B}$ term results in an additional change in the distribution function, given by $\Bigl( e E v_{x}(\bm{p}) + \frac{e^{2}}{c} (E B) \boldsymbol{\Omega}_{\boldsymbol{p}}^{\chi} \cdot \bm{v}_{\bm{p}} \Bigr)$. In addition, the $\bm{E}\cdot\bm{B}$ term is also responsible for the transverse deflection, forbidden as long as only the Lorentz force and Berry curvature are concerned. We emphasize that the topological $\bm{E}\cdot\bm{B}$ term is beyond the contribution of the Berry curvature only. In other words, such a term will not arise in the graphene structure.

\subsubsection{Longitudinal magneto-conductivity}

It is straightforward to find the ``longitudinal" magneto-conductivity, given by
\bqa && \sigma_{xx}^{\chi} = e^{2} \int \frac{d^{3} \bm{p}}{(2\pi)^{3}} \Bigl( - \frac{\partial }{\partial \epsilon} f_{eq} (\epsilon) \Bigr) \Bigl( 1 + \frac{e B}{c} \Omega_{x}^{\chi}(\boldsymbol{p}) \Bigr)^{-1} \Bigl\{ v_{x}(\boldsymbol{p}) + \frac{e B}{c} (\boldsymbol{\Omega}_{\boldsymbol{p}}^{\chi} \cdot \boldsymbol{v}_{\boldsymbol{p}}) \Bigr\}^{2} \frac{1}{\Gamma_{imp} + \alpha} \nn && + e^{2} \int \frac{d^{3} \bm{p}}{(2\pi)^{3}} \Bigl( - \frac{\partial }{\partial \epsilon} f_{eq} (\epsilon) \Bigr) \Bigl( 1 + \frac{e B}{c} \Omega_{x}^{-\chi}(\bm{p}) \Bigr)^{-1} \Bigl\{ v_{x}(\boldsymbol{p}) + \frac{e B}{c} (\boldsymbol{\Omega}_{\boldsymbol{p}}^{\chi} \cdot \boldsymbol{v}_{\boldsymbol{p}}) \Bigr\} \Bigl\{ v_{x}(\boldsymbol{p}) + \frac{e B}{c} \Omega_{x}^{-\chi}(\bm{p}) v_{x}(\boldsymbol{p}) \Bigr\} \frac{\Gamma_{imp}'}{[\Gamma_{imp} + \alpha]^{2}} \nn && - e^{2} \Bigl( \frac{e B}{c} \Bigr)^{2}\int \frac{d^{3} \bm{p}}{(2\pi)^{3}} \Bigl( - \frac{\partial}{\partial \epsilon} f_{eq} (\epsilon) \Bigr) \Bigl( 1 + \frac{e B}{c} \Omega_{x}^{\chi}(\boldsymbol{p}) \Bigr)^{-1} (\boldsymbol{\Omega}_{\boldsymbol{p}}^{\chi} \cdot \boldsymbol{v}_{\boldsymbol{p}}) \nn && \Bigl\{ v_{y}({\bm{p}}) \frac{1}{\Gamma_{imp} + \alpha} \frac{ [\Omega_{c}^{\chi}(\bm{p})]^{2} \Omega_{y}^{\chi}(\bm{p}) - (\Gamma_{imp} + \alpha) \Omega_{c}^{\chi}(\bm{p}) \Omega_{z}^{\chi}(\bm{p})}{ [\Gamma_{imp} + \alpha]^{2} + [ \Omega_{c}^{\chi}(\bm{p})]^{2} } + v_{z}({\bm{p}}) \frac{1}{\Gamma_{imp} + \alpha} \frac{ (\Gamma_{imp} + \alpha) \Omega_{c}^{\chi}(\bm{p}) \Omega_{y}^{\chi}(\bm{p}) + [\Omega_{c}^{\chi}(\bm{p})]^{2} \Omega_{z}^{\chi}(\bm{p})}{ [\Gamma_{imp} + \alpha]^{2} + [ \Omega_{c}^{\chi}(\bm{p})]^{2} } \Bigr\} \nn && - e^{2} \Bigl( \frac{e B}{c} \Bigr)^{2}\int \frac{d^{3} \bm{p}}{(2\pi)^{3}} \Bigl( - \frac{\partial}{\partial \epsilon} f_{eq} (\epsilon) \Bigr) \Bigl( 1 + \frac{e B}{c} \Omega_{x}^{-\chi}(\boldsymbol{p}) \Bigr)^{-1} (\boldsymbol{\Omega}_{\boldsymbol{p}}^{\chi} \cdot \boldsymbol{v}_{\boldsymbol{p}}) \nn && \Bigl\{ - v_{y}({\bm{p}}) \frac{\Gamma_{imp}'}{\Gamma_{imp} + \alpha} \frac{(\Gamma_{imp} + \alpha) \Omega_{y}^{-\chi}(\bm{p}) + \Omega_{c}^{\chi}(\bm{p}) \Omega_{z}^{-\chi}(\bm{p})}{ [\Gamma_{imp} + \alpha]^{2} + [ \Omega_{c}^{\chi}(\bm{p})]^{2} } + v_{z}({\bm{p}}) \frac{\Gamma_{imp}'}{\Gamma_{imp} + \alpha} \frac{\Omega_{c}^{\chi}(\bm{p}) \Omega_{y}^{-\chi}(\bm{p}) - (\Gamma_{imp} + \alpha) \Omega_{z}^{-\chi}(\bm{p})}{ [\Gamma_{imp} + \alpha]^{2} + [ \Omega_{c}^{\chi}(\bm{p})]^{2} } \Bigr\} . \nn \eqa

Taking square-dependent terms for both the velocity and the Berry curvature as the leading order, we simplify these formulae as follows
\bqa && \sigma_{xx}^{\chi} \approx e^{2} \int \frac{d^{3} \bm{p}}{(2\pi)^{3}} \Bigl( - \frac{\partial }{\partial \epsilon} f_{eq} (\epsilon) \Bigr) \Bigl( 1 + \frac{e B}{c} \Omega_{x}^{\chi}(\boldsymbol{p}) \Bigr)^{-1} \Bigl\{ [v_{x}(\boldsymbol{p})]^{2} + \Bigl(\frac{e B}{c}\Bigr)^{2} \Bigl( [\Omega_{x}^{\chi}(\boldsymbol{p})]^{2} [v_{x}(\boldsymbol{p})]^{2} + [\Omega_{y}^{\chi}(\boldsymbol{p})]^{2} [v_{y}(\boldsymbol{p})]^{2} \nn && + [\Omega_{z}^{\chi}(\boldsymbol{p})]^{2} [v_{z}(\boldsymbol{p})]^{2} \Bigr) \Bigr\} \frac{1}{\Gamma_{imp} + \alpha} + e^{2} \int \frac{d^{3} \bm{p}}{(2\pi)^{3}} \Bigl( - \frac{\partial }{\partial \epsilon} f_{eq} (\epsilon) \Bigr) [v_{x}(\boldsymbol{p})]^{2} \frac{\Gamma_{imp}'}{[\Gamma_{imp} + \alpha]^{2}} \nn && - e^{2} \Bigl( \frac{e B}{c} \Bigr)^{2}\int \frac{d^{3} \bm{p}}{(2\pi)^{3}} \Bigl( - \frac{\partial}{\partial \epsilon} f_{eq} (\epsilon) \Bigr) \Bigl( 1 + \frac{e B}{c} \Omega_{x}^{\chi}(\boldsymbol{p}) \Bigr)^{-1} \Bigl( [\Omega_{y}^{\chi}(\bm{p})]^{2} [v_{y}({\bm{p}})]^{2} + [\Omega_{z}^{\chi}(\bm{p})]^{2} [v_{z}({\bm{p}})]^{2} \Bigr) \frac{1}{\Gamma_{imp} + \alpha} \frac{ \omega_{c}^{2}}{ [\Gamma_{imp} + \alpha]^{2} + \omega_{c}^{2} } \nn && + e^{2} \Bigl( \frac{e B}{c} \Bigr)^{2}\int \frac{d^{3} \bm{p}}{(2\pi)^{3}} \Bigl( - \frac{\partial}{\partial \epsilon} f_{eq} (\epsilon) \Bigr) \Bigl( 1 + \frac{e B}{c} \Omega_{x}^{-\chi}(\boldsymbol{p}) \Bigr)^{-1} \Bigl( \Omega_{y}^{\chi}(\bm{p}) \Omega_{y}^{-\chi}(\bm{p}) [v_{y}({\bm{p}})]^{2} + \Omega_{z}^{\chi}(\bm{p}) \Omega_{z}^{-\chi}(\bm{p}) [v_{z}({\bm{p}})]^{2} \Bigr) \frac{\Gamma_{imp}'}{ [\Gamma_{imp} + \alpha]^{2} + \omega_{c}^{2}} . \nn \eqa

Performing the momentum integral and summing contributions of both chiralities with $\bm{\Omega}_{\bm{p}}^{+} \approx - \bm{\Omega}_{\bm{p}}^{-} = \bm{\Omega}_{\bm{p}}$, we reach the following expression
\bqa && \sigma_{xx} = 2 \sigma \Bigl\{ 1 + \mathcal{C}_{ABJ} \Bigl(\frac{e B}{c}\Bigr)^{2} + \frac{\Gamma_{imp}'}{\Gamma_{imp}} \frac{1}{1 + \alpha/\Gamma_{imp}} \Bigr\} \frac{1}{1 + \alpha/\Gamma_{imp}} \nn && - \frac{4}{3} \sigma \mathcal{C}_{ABJ} m^{2} \omega_{c}^{2} \Bigl( \frac{\omega_{c}^{2} / \Gamma_{imp}^{2}}{1 + \alpha / \Gamma_{imp}} + \frac{\Gamma_{imp}'}{\Gamma_{imp}} \Bigr) \frac{ 1 }{ [1 + \alpha / \Gamma_{imp}]^{2} + \omega_{c}^{2} / \Gamma_{imp}^{2} } , \eqa where undefined conductivities are given by
\bqa && \sigma \approx \frac{e^{2}}{\Gamma_{imp}} \int \frac{d^{3} \bm{p}}{(2\pi)^{3}} \Bigl( - \frac{\partial }{\partial \epsilon} f_{eq} (\epsilon) \Bigr) \frac{|\bm{v}_{\bm{p}}|^{2}}{3} , ~~~~~ \sigma \mathcal{C}_{ABJ} \approx \frac{e^{2}}{\Gamma_{imp}} \int \frac{d^{3} \bm{p}}{(2\pi)^{3}} \Bigl( - \frac{\partial }{\partial \epsilon} f_{eq} (\epsilon) \Bigr) \frac{|\bm{v}_{\bm{p}}|^{2}}{3} |\bm{\Omega}_{\bm{p}}|^{2} . \eqa In order to simplify the expression, we assumed a simple Fermi surface, given by $[v_{x}(\bm{p})]^{2} = [v_{y}(\bm{p})]^{2} = [v_{z}(\bm{p})]^{2} = \frac{|\bm{v}_{\bm{p}}|^{2}}{3}$ and $[\Omega_{x}(\bm{p})]^{2} = [\Omega_{y}(\bm{p})]^{2} = [\Omega_{z}(\bm{p})]^{2} = \frac{|\bm{\Omega}_{\bm{p}}|^{2}}{3}$. If we take the limit of $\Gamma_{imp}' / \Gamma_{imp} \rightarrow 0$, this expression is further simplified as \bqa && \sigma_{xx} = 2 \sigma \Bigl\{ 1 + \mathcal{C}_{ABJ} \Bigl(\frac{e B}{c}\Bigr)^{2} \Bigr\} \frac{1}{1 + \alpha/\Gamma_{imp}} - \frac{4}{3} \sigma \mathcal{C}_{ABJ} m^{2} \omega_{c}^{2} \frac{1}{1 + \alpha / \Gamma_{imp}} \frac{ \omega_{c}^{2} / \Gamma_{imp}^{2} }{ [1 + \alpha / \Gamma_{imp}]^{2} + \omega_{c}^{2} / \Gamma_{imp}^{2} } . \nonumber \eqa Focusing on the low-field region, we obtain \bqa && \sigma_{xx} = 2 \sigma \Bigl\{ 1 + \mathcal{C}_{ABJ} \Bigl(\frac{e B}{c}\Bigr)^{2} \Bigr\} \frac{1}{1 + \alpha/\Gamma_{imp}} , \nonumber \eqa referred to as the ``positive" magneto-conductivity, where the $B^{2}$ contribution results from the $\bm{E}\cdot\bm{B}$ term. Inserting the weak anti-localization correction into the above formula and considering $l_{ph}^{-1} = (\mathcal{C}'/\mathcal{C}) \sqrt{B}$ with a positive constant $\mathcal{C}'$, we find \bqa && \sigma_{xx} = \frac{2}{\rho_{imp}} \Bigl\{ 1 + \mathcal{C}_{ABJ} \Bigl(\frac{e B}{c}\Bigr)^{2} \Bigr\} \frac{1}{1 - \mathcal{C} e^{2} N_{F} \rho_{imp} l_{imp}^{-1} + \mathcal{C}' e^{2} N_{F} \rho_{imp} \sqrt{B}} , \label{LMR_ABJ} \eqa which turns out to fit the experimental data well \cite{Negative_LMR3}.

In order to explain the experimental data of Ref. \cite{Negative_LMR3}, we introduced two contributions for magneto-conductivity, where one results from Weyl electrons near the $\bm{L}-$point of the momentum space and the other comes from normal electrons near the $\bm{T}-$point. Subtracting out the cyclotron contribution of normal electrons in the transverse setup ($\bm{B} \perp \bm{E}$), we could fit the data based on the three-dimensional weak-antilocalization formula, given by Weyl electrons, where the weak-antilocalization correction has been Taylor-expanded for the weak-field region below $1.2$ T. On the other hand, the cyclotron contribution around the $\bm{T}-$point almost vanishes for the longitudinal setup ($\bm{B} \parallel \bm{E}$) as it must be, and the residual resistivity for normal electrons is almost identical with that of the transverse setup. Subtracting out the $\bm{T}-$point contribution, we could fit the data with Eq. (\ref{LMR_ABJ}) in the regime of the weak magnetic field below $1.2$ T, where the weak-antilocalization correction has been also Taylor-expanded. Again, the weak-antilocalization correction turns out to be almost identical with that of the transverse setup while we have an additional constant $\mathcal{C}_{ABJ}$ in the longitudinal setup, the origin of which is the chiral anomaly.

\subsubsection{Hall conductivity}

Following the same strategy as that of the magneto-conductivity, it is straightforward to find the Hall conductivity around each Weyl point, given by
\bqa && \sigma_{yx}^{\chi} = - e^{2} \int \frac{d^{3} \bm{p}}{(2\pi)^{3}} \Omega_{z}^{\chi}(\bm{p}) f_{eq}(\bm{p}) \nn && + e^{2} \Bigl( \frac{e B}{c} \Bigr) \int \frac{d^{3} \bm{p}}{(2\pi)^{3}} \Bigl( - \frac{\partial }{\partial \epsilon} f_{eq} (\epsilon) \Bigr) [v_{y}(\bm{p})]^{2} \Bigl( 1 + \frac{e B}{c} \Omega_{x}^{\chi}(\bm{p}) \Bigr)^{-1} \Omega_{y}^{\chi}(\bm{p}) \frac{1}{\Gamma_{imp} + \alpha} \nn && - e^{2} \Bigl( \frac{e B}{c} \Bigr) \int \frac{d^{3} \bm{p}}{(2\pi)^{3}} \Bigl( - \frac{\partial }{\partial \epsilon} f_{eq} (\epsilon) \Bigr) [v_{y}(\bm{p})]^{2} \Bigl( 1 + \frac{e B}{c} \Omega_{x}^{\chi}(\bm{p}) \Bigr)^{-1} \frac{1}{\Gamma_{imp} + \alpha} \frac{ [\Omega_{c}^{\chi}(\bm{p})]^{2} \Omega_{y}^{\chi}(\bm{p}) - (\Gamma_{imp} + \alpha) \Omega_{c}^{\chi}(\bm{p}) \Omega_{z}^{\chi}(\bm{p})}{ [\Gamma_{imp} + \alpha]^{2} + [ \Omega_{c}^{\chi}(\bm{p})]^{2} } \nn && + e^{2} \Bigl( \frac{e B}{c} \Bigr) \int \frac{d^{3} \bm{p}}{(2\pi)^{3}} \Bigl( - \frac{\partial }{\partial \epsilon} f_{eq} (\epsilon) \Bigr) [v_{y}(\bm{p})]^{2} \Bigl( 1 + \frac{e B}{c} \Omega_{x}^{-\chi}(\bm{p}) \Bigr)^{-1}  \frac{\Gamma_{imp}'}{\Gamma_{imp} + \alpha} \frac{(\Gamma_{imp} + \alpha) \Omega_{y}^{-\chi}(\bm{p}) + \Omega_{c}^{\chi}(\bm{p}) \Omega_{z}^{-\chi}(\bm{p})}{ [\Gamma_{imp} + \alpha]^{2} + [ \Omega_{c}^{\chi}(\bm{p})]^{2} } . \eqa Here, we keep only $[v_{y}(\bm{p})]^{2}$-dependent terms except for the Berry-curvature term, consistent with the strategy of the case of the normal setup. The first term is the anomalous Hall effect resulting from the Berry curvature while all other terms are another type of the anomalous Hall effect originating from the chiral anomaly, where the topological $\bm{E}\cdot\bm{B}$ term gives rise to an additional force beyond the conventional Lorentz force. However, we find that the anomaly-induced anomalous Hall effect does not exist, inserting $\bm{\Omega}_{\bm{p}}^{\chi} \propto \chi \frac{\bm{\hat{p}}}{|\bm{p} - \chi g_{\psi} \bm{B}|^{2}}$ with $\bm{B} = B \bm{\hat{x}}$ into the above expression and performing the momentum integration. In other words, we have $\sigma_{yx}^{\chi} = 0$.

\section{Perspectives}

The Boltzmann-equation approach describes anomalous transport phenomena of Weyl metal such as chiral magnetic effect and negative magneto-resistivity quite successfully, where the topological structure of Weyl metal can be introduced via the semi-classical equation-of-motion approach with Berry curvature. However, we believe that our microscopic understanding on these phenomena is incomplete in the respect that we do not know how to evaluate such transport coefficients based on the diagrammatic approach. For example, we speculate that a conventional diagrammatic approach will not allow the $B^{2}$ contribution in the longitudinal magneto-conductivity, giving rise to only the Drude part (with weak anti-localization corrections). First of all, an effective field theory has not been proposed yet, which must incorporate both the Berry curvature and chiral anomaly \cite{TLFL_EFT}. The chiral anomaly have to be introduced explicitly into the effective field theory as a local curvature term because such a term is not purely topological any more as the case of axion electrodynamics \cite{Axion_EM}. Of course, this effective field theory must reproduce essentially the same Boltzmann-equation framework investigated in the present paper. In addition, both the chiral magnetic effect and negative magneto-resistivity should be recovered within the conventional diagrammatic approach, based on this effective field theory. We expect that this theoretical framework takes the first step toward ``topological" Landau Fermi liquid theory, incorporating both electron correlations and topological aspects.

\section*{Acknowledgement}

This study was supported by the Ministry of Education, Science, and Technology (No. 2012R1A1B3000550 and No. 2011-0030785) of the National Research Foundation of Korea (NRF) and by TJ Park Science Fellowship of the POSCO TJ Park Foundation. K.-S. Kim appreciates fruitful discussions with Chushun Tian.


\begin{thebibliography}{9}
\bibitem{Kane_Mele_Model} C. L. Kane and E. J. Mele, Phys. Rev. Lett. {\bf 95}, 146802 (2005); C. L. Kane and E. J. Mele, Phys. Rev. Lett. {\bf 95}, 226801 (2005).
\bibitem{Zhang_Bernevig_Model} B. Andrei Bernevig, Taylor L. Hughes, and Shou-Cheng Zhang, Science {\bf 314}, 1757 (2006).
\bibitem{3D_TI_Z2index_FuKane} L. Fu, C. L. Kane, and E. J. Mele, Phys. Rev. Lett. {\bf 98}, 106803 (2007).
\bibitem{3D_TI_Z2index_Balents} J. E. Moore and L. Balents, Phys. Rev. B {\bf 75}, 121306 (2007).
\bibitem{3D_TI_Z2index_Roy} R. Roy, Phys. Rev. B {\bf 79}, 195322 (2009); R. Roy, Phys. Rev. B {\bf 79}, 195321 (2009).
\bibitem{Chiral_Anomaly} S. Adler, Phys. Rev. {\bf 177}, 2426 (1969); J. S. Bell and R. Jackiw, Nuovo Cimento A {\bf 60}, 4 (1969).
\bibitem{Chiral_Magnetic_Effect1} Dam Thanh Son and Naoki Yamamoto, Phys. Rev. Lett. {\bf 109}, 181602 (2012).
\bibitem{Chiral_Magnetic_Effect2} M. A. Stephanov and Y. Yin, Phys. Rev. Lett. {\bf 109}, 162001 (2012).
\bibitem{CME_Kubo1} K. Landsteiner, E. Megias, and F. Pena-Benitez, Phys. Rev. Lett. {\bf 107}, 021601 (2011).
\bibitem{CME_Kubo2} Y. Chen, Si Wu, and A. A. Burkov, Phys. Rev. B {\bf 88}, 125105 (2013).
\bibitem{Chiral_Magnetic_Effect3} Kenji Fukushima, Dmitri E. Kharzeev, and Harmen J. Warringa, Phys. Rev. D {\bf 78}, 074033 (2008).
\bibitem{Chiral_Magnetic_Effect4} Gokce Basar, Dmitri E. Kharzeev, and Ho-Ung Yee, Phys. Rev. B {\bf 89}, 035142 (2014).
\bibitem{Negative_LMR1} H. B. Nielsen and M. Ninomiya, Phys. Lett. {\bf 130B}, 389 (1983).
\bibitem{Negative_LMR2} D. T. Son and B. Z. Spivak, Phys. Rev. B {\bf 88}, 104412 (2013).
\bibitem{Negative_LMR3} Heon-Jung Kim, Ki-Seok Kim, J.-F. Wang, M. Sasaki, N. Satoh, A. Ohnishi, M. Kitaura, M. Yang, and L. Li, Phys. Rev. Lett. {\bf 111}, 246603 (2013).
\bibitem{Haldane} F. D. M. Haldane, Phys. Rev. Lett. {\bf 93}, 206602 (2004).
\bibitem{TI_Band_Structure} Gil Young Cho, arXiv:1110.1939v2.
\bibitem{BiSb1} L. Fu and C. L. Kane, Phys. Rev. B {\bf 76}, 045302 (2007).
\bibitem{BiSb2} J. C.Y. Teo, L. Fu, and C. L. Kane, Phys. Rev. B {\bf 78}, 045426 (2008).
\bibitem{BiSb3} H. Guo, K. Sugawara, A. Takayama, S. Souma, T. Sato, N. Satoh, A. Ohnishi, M. Kitaura, M. Sasaki, Q.-K. Xue, and T. Takahashi, Phys. Rev. B {\bf 83}, 201104(R) (2011).
\bibitem{Weyl_Metal1} S. Murakami, New J. Phys. {\bf 9}, 356 (2007).
\bibitem{Weyl_Metal2} A. A. Burkov and L. Balents, Phys. Rev. Lett. {\bf 107}, 127205 (2011).
\bibitem{Weyl_Metal3} Pavan Hosur and Xiaoliang Qi, arXiv:1309.4464v1.
\bibitem{Review_Graphene} A. H. Castro Neto, F. Guinea, N. M. R. Peres, K. S. Novoselov, and A. K. Geim, Rev. Mod. Phys. {\bf 81}, 109 (2009).
\bibitem{NoGoTheorem1} H. B. Nielsen and M. Ninomiya, Nucl. Phys. B {\bf 193}, 173 (1981).
\bibitem{NoGoTheorem2} H. B. Nielsen and M. Ninomiya, Phys. Lett. B {\bf 105}, 219 (1981).
\bibitem{Peskin_Schroeder} M. E. Peskin and D. V. Schroeder, {\it An Introduction to Quantum Field Theory} (Addison-Wesley Publishing Company, New York, 1995).
\bibitem{Jho_Kim} Y.-S. Jho and K.-S. Kim, Phys. Rev. B {\bf 87}, 205133 (2013).
\bibitem{Semiclassical_Eqs1} D. Xiao, M.-C. Chang, and Q. Niu, Rev. Mod. Phys. {\bf 82}, 1959 (2010).
\bibitem{Semiclassical_Eqs2} N. Nagaosa, J. Sinova, S. Onoda, A. H. MacDonald, and N. P. Ong, Rev. Mod. Phys. {\bf 82}, 1539 (2010).
\bibitem{WAL_Boltzmann} Selman Hershfield and Vinay Ambegaokar, Phys. Rev. B {\bf 34}, 2147 (1986).
\bibitem{AHE_Kubo1} P. Goswami and Sumanta Tewari, Phys. Rev. B {\bf 88}, 245107 (2013).
\bibitem{AHE_Kubo2} A. A. Zyuzin and A. A. Burkov, Phys. Rev. B {\bf 86}, 115133 (2012); Y. Chen, D. L. Bergman, and A. A. Burkov, Phys. Rev. B {\bf 88}, 125110 (2013).
\bibitem{Mahan_Book} G. D. Mahan, Many-Particle Physics, 3rd ed. (Kluwer Academic/Plenum, New York, 2000).
\bibitem{QBE_WM1} Jian-Hua Gao, Zuo-Tang Liang, Shi Pu, Qun Wang, and Xin-Nian Wang, Phys. Rev. Lett. {\bf 109}, 232301 (2012).
\bibitem{QBE_WM2} Jiunn-Wei Chen, Shi Pu, Qun Wang, and Xin-Nian Wang, Phys. Rev. Lett. {\bf 110}, 262301 (2013).
\bibitem{QBE_WM3} Dam Thanh Son and Naoki Yamamoto, Phys. Rev. D {\bf 87}, 085016 (2013).
\bibitem{Review_Disorder} Patrick A. Lee and T. V. Ramakrishnan, Rev. Mod. Phys. {\bf 57}, 287 (1985).
\bibitem{TLFL_EFT} Recently, an effective field theory has been proposed in the first quantization form, where the mathematical expression for the Berry curvature is rewritten as a gauged Wess-Zumino-Witten (WZW) term in five dimensional space with one auxiliary space. The five dimensional Chern-Simons term in the gauged Wess-Zumino-Witten (WZW) term appears to recover the chiral anomaly in four dimensional space. See Ismail Zahed, Phys. Rev. Lett. {\bf 109}, 091603 (2012) and Gokce Basar, Dmitri E. Kharzeev, and Ismail Zahed, Phys. Rev. Lett. {\bf 111}, 161601 (2013).
\bibitem{Axion_EM} F. Wilczek, Phys. Rev. Lett. {\bf 58}, 1799 (1987).
\end{thebibliography}
\end{document}